 \documentclass[manuscript]{aastex62}
\shorttitle{2016 outburst of H~1743--322}
\shortauthors{Chand et al.}

\usepackage{threeparttable}
\usepackage{url}

\begin{document}

\title{2016 outburst of H~1743--322: \textit{XMM-Newton} and \textit{NuSTAR} view}

\author{Swadesh Chand}
\affiliation{Department of Pure and Applied Physics, Guru Ghasidas Vishwavidyalaya (A Central University), Bilaspur (C. G)- 495009, India}

\author{V. K. Agrawal}
\affiliation{Space Astronomy Group, ISITE Campus, ISRO Satellite Centre, Bangalore - 560037, India}

\author{G. C. Dewangan}
\affiliation{Inter-University Centre for Astronomy and Astrophysics, Post Bag 4, Ganeshkhind, Pune - 411007, India}

\author{Prakash Tripathi}
\affiliation{Inter-University Centre for Astronomy and Astrophysics, Post Bag 4, Ganeshkhind, Pune - 411007, India}

\author{Parijat Thakur}
\affiliation{Department of Pure and Applied Physics, Guru Ghasidas Vishwavidyalaya (A Central University), Bilaspur (C. G)- 495009, India}
\nocollaboration

\correspondingauthor{Parijat Thakur}
\email{parijat@associates.iucaa.in, parijatthakur@yahoo.com}



\begin{abstract}
We report the detection of type C QPO along with the upper harmonic in the commensurate ratio of 1:2 in the two observations of the low-mass black hole transient H~1743--322 jointly observed by \textit{XMM-Newton} and \textit{NuSTAR} during the 2016 outburst. We find that the QPO and the upper harmonic exhibit shifts in their centroid frequencies in the second observation with respect to the first one. The hardness intensity diagram implies that in contrast to 2008 and 2014 failed outbursts, 2016 outburst was a successful one. We also detect the presence of a broad iron K$\alpha$ line at $\sim$6.5 keV and reflection hump in the energy range of 15--30 keV in both the observations. Along with the shape of the power  density spectra, the nature of the characteristic frequencies and the fractional rms amplitude of the timing features imply that the source stayed in the low/hard state during these observations. Moreover, photon index and other spectral parameters also indicate the low/hard state behavior of the source. Unlike the soft lag detected in this source during the 2008 and 2014 failed outbursts, we observe hard time-lag of $0.40\pm0.15$ and $0.32\pm0.07$ s in the 0.07--0.4 Hz frequency range in the two observations during the 2016 outburst. The correlation between the photon index and the centroid frequency of the QPO is consistent with the previous results. Furthermore, the high value of the Comptonized fraction and the weak thermal component indicate that the QPO is being modulated by the Comptonization process.

\end{abstract}



\section{Introduction} \label{sec:intro}

Low-mass black hole X-ray binaries consist of a low-mass companion star
($\lesssim 1$ \(\textup{M}_\odot\)) gravitationally bound to a stellar
mass black hole \citep{Steiner et al. 2012}. The companion star feeds
material to the black hole via Roche lobe overflow, resulting in the
formation of an accretion disc. The viscous forces between the different
layers of the accretion disc near the black hole raise the temperature
upto $10^7$ K, and the source primarily emits X-rays \citep{Steiner et al. 2012, Motta
et al. 2017}. Most of the black hole X-ray binaries
(hereafter BHXRBs) are known to be transient that stay in quiescence for a long time and show
outbursts very sporadically. These outbursts can last
from several days to months during which the luminosity of the source
is increased by several orders of magnitude \citep{Tanaka and Shibazaki
1996, Shidatsu et al. 2014, Plant et al. 2015}. The BHXRBs can undergo generally four states during an outburst, namely, the low/hard state
(LHS), the hard-intermediate state (HIMS), the soft-intermediate state
(SIMS) and the high/soft state (HSS) \citep{Belloni et al. 2005}. The classification of the states relies upon the 
detailed spectral and timing behavior of the source during an outburst.

The spectrum of the LHS is dominated by a hard power-law with photon index
$<$2 and a high-energy cutoff $\approx$ 100 keV \citep{Motta et al. 2009,
Shidatsu et al. 2014}. The power-law component is thought to originate
from the Compton up-scattering of the soft X-ray photons from the disc
by the hot electrons in corona. The power density spectra (hereafter
PDS) in the LHS state show strong variability with fractional rms $\sim$  30\%
\citep{McClintock and Remillard 2006, Belloni et al. 2011, Zhou et
al. 2013, Shidatsu et al. 2014, Ingram et al. 2017}. The transition of
the BHXRBs to the HSS occurs via the two intermediate states
(i.e., HIMS and SIMS). However, the transition from the hard to soft is not a smooth one 
and in many outbursts, several excursions to harder and softer states have been observed.
As the source moves from the hard state to the soft state, the power-law component starts to steepen 
(upto $\Gamma \sim 2.5$) and the X-ray continuum becomes increasingly
dominated by emission from an optically thick and geometrically thin
accretion disc with a few percent of fractional rms variability 
\citep{Shakura and Sunyaev 1973, Belloni et al. 2011}.
In the soft state, the accretion disc either reaches closer to the  inner most stable circular orbit (hereafter ISCO)
 or extends down to the ISCO \citep{Gierlinski and Done 2004, Steiner
et al. 2010}. It is worth mentioning here that there is an ongoing debate on the nature of extent of the
accretion disc in the LHS. However, several workers has found that a hot advection dominated
accretion flow (ADAF), as proposed by \citet{Esin et al. 1997}, replaces
the geometrically thin and optically thick accretion disc,
 introducing a truncated inner disc \citep{McClintock et al. 1995, Narayan and Yi 1995, Narayan et al. 1996, Esin et al. 2001, McClintock et al. 2001, McClintock et al. 2003, Plant et al. 2015}.

Another salient observational feature of BHXRBs is the X-ray reflection. This appears when the  hard Comptonized X-rays from the
corona gets reflected from the disc giving rise to a reflection
hump in the  $\sim$10--40 keV and a fluorescent iron line $K_{\alpha}$ line at $\sim$ 6.4--6.9 keV. The iron line may be distorted due to special and 
general relativistic effects \citep{Fabian et al. 1989, Reynolds and Nowak 2003, Miller 2007, Ingram et al. 2017}. Modeling of the broad iron line with relativistic reflection gives an alternative way to measure
the inner disc radius. 
The modeling of the reflection
continuum can also play a crucial role in understanding the inner accretion
dynamics of the disc, as well as to probe the key parameters like the
black hole spin and the disc inclination. It is worth mentioning here that the relativistic reflection model RELXILL is being widely used to model angle-dependent X-ray reflection and allows to estimate parameters such as the  inner disc radius, black hole spin, disc inclination, iron abundance and reflection fraction \citep{Dauser et al. 2010, Garcia and Kallman 2010, Garcia et al. 2011, Garcia et al. 2013, Dauser et al. 2013, Dauser et al. 2014, Garcia et al. 2014}.

Quasi-periodic oscillations (QPOs), which appear as broad peaks, are often observed in the
X-ray emission from the black hole transients (hereafter BHTs). QPOs are categorized into the following three types: (i) high-frequency QPOs (HFQPOs) ($\sim$30--450 Hz) (ii) low-frequency QPOs (LFQPOs) ($\sim$0.05--30 Hz) and (iii) very low-frequency QPOs ($\sim$mHz)\citep{Morgan et al. 1997, Belloni et al. 2000, Trudolyubov et al. 2001,  Casella et al. 2005, Motta et al. 2011, Altamirano et al. 2011, Altamirano et al. 2012, Belloni et al. 2012, Alam et al. 2014, AgrawalNandi2015}. Depending upon a few parameters such as  quality factor, ($Q$), and shape of the PDS continuum, low-frequency QPOs are classified into type A, B and C. 
 Type A QPOs appear as broad peaks around $\sim$6--8 Hz with few percent rms, whereas the type B QPOs show stronger rms (upto $\sim4\%$) compared to the type A QPOs. On the other hand, type C QPO appears as a narrow and variable peak with strong fractional rms $\geq10\%$ \citep{Casella et al. 2004, Motta et al. 2011}. Though the exact mechanism for the origin of the QPOs is  still not clear, it has been found in earlier studies that centroid frequencies of QPOs are correlated with the spectral index, as well as the disc flux \citep{Sobczak et al. 2000, Titarchuk and Fiorito 2004, Shaposhnikov and Titarchuk 2007}, indicating the coupling between QPOs and the structure of the inner disc \citep{Shidatsu et al. 2014}.
 
Strong variability of BHXRBs may also be related to the the time-lags found between the lightcurves in different energy bands. \citet{Priedhorsky1979} and \citet{Nolan1981} first found the presence of time-lags in Cygnus X-1, as well as several other BHXRBs. Many BHXRBs show hard X-ray lags, where the hard photons are found to be delayed with respect to the soft ones \citep{Page1981, Miyamoto1988, Nowak1999a, Nowak1999b, Grinberg2014}. These hard X-ray lags are generally thought to be originated due to the propagation of mass accretion rate fluctuation in the accretion disc and are of the magnitude of 1 percent of the variability time scale \citep{DeMarco2013, ArevaloandUttley2006}. On the other hand, the soft X-ray lags, where the soft X-ray photons lag the hard ones, are caused due to the reflection of the coronal X-ray irradiation by the accretion disc. This time delay is named as reverberation lag and can provide important information about the geometry of the innermost region of the BHXRBs \citep{DeMarco2013, MarcoPonti2016, DeMarco2017, KaraErin2019}.
 
The low-mass BHT H~1743--322 was discovered in 1977 using \textit{Ariel-V} \citep{Kaluzienski and Holt 1977} and is located at a distance of 8.5$\pm$0.8 kpc \citep{Steiner et al. 2012}. This source is well known for its 
transient nature and has shown frequent outbursts with an average interval of $\sim$200 days \citep{Shidatsu et al. 2012,
 Shidatsu et al. 2014}. After a prolonged gap from its discovery, the brightest outburst of 
 H~1743--322 in 2003 was 
 detected by \textit{INTEGRAL} \citep{Revnivstev 2003} and \textit{RXTE} \citep{Markwardt and Swank 2003}. 
 \textit{RXTE} observation of this outburst resulted in the detection of a pair of HFQPOs at 240 and
  160 Hz \citep{Homan et al. 2005, Remillard et al. 2006}. Similar timing signature has also been found in a few
   other dynamical BHXRBs \citep{McClintock and Remillard 2006, McClintock et al. 2009}. Using the \textit{VLA} 
   observation of 2003 outburst and by applying a symmetric kinematic model for the jet trajectories, \citet{Steiner et al. 2012} 
   estimated the source distance and inclination angle to be 8.5 $\pm$0.8 kpc and 75$^\circ\pm3^\circ$, respectively.
 They also found the spin parameter to be 0.2$\pm$ 0.3 using the \textit{RXTE} observation of the 2003 outburst. 
 In addition to this, \citet{srivivrao2009} found QPOs with a truncated accretion disc using the \textit{RXTE} 
 observations of 2003 outburst, when the source was in the steep powerlaw state. Before the October 2008 outburst,
  a few additional outbursts were detected which could not be studied extensively due to lack of sufficient observations.  
However, the October 2008 outburst was termed as a `failed outburst' as the source never reached the HSS due to sudden decrease in the mass accretion rate \citep{Capitanio et al. 2009}. Another 
 outburst in July 2009 was detected by \textit{Swift/BAT} telescope \citep{Krimm et al. 2009, Motta et al. 2010}, which
 was followed by three more outbursts detected by \textit{RXTE} in December 2009, August 2010 and April 2011 \citep{Zhou et al. 2013}.
  Using these 2010 and 2011 outbursts, \citet{Molla et al. 2017} estimated the mass 
  of the black hole to be $11.21^{+1.65}_{-1.96}$ M$_\odot$ by combining the two methods: using the Two Component Advective Flow (TCAF) model and the correlation between the photon index versus QPO frequency \citep{DewanganTitarchukandGriffiths2006}.
  Apart from the above, few successive outbursts were reported in December 2011, 
   January 2012 \citep{Negoro et al. 2012}, September 2012 \citep{Shidatsu et al. 2012, Shidatsu et al. 2014}, 
   and August 2013 \citep{Nakahira et al. 2013}. Following these outbursts, another outburst took place in 2014, 
   which was observed quasi-simultaneously by both \textit{XMM-Newton} and \textit{NuSTAR}. 
   Using the \textit{Swift/XRT} monitoring of the 2014 outburst, \citet{Stiele and Yu 2016} reported the 2014 outburst 
   as a failed one as the source never reached the HSS during the entire outburst. In addition to this, using the \textit{XMM-Newton} 
   observation of this outburst, the authors reported a low-frequency QPO and an upper harmonic 
   at $\sim$0.25 and $\sim$0.51 Hz, respectively.   
   Moreover, \citet{Ingram et al. 2017} used both the \textit{XMM-Newton}, as well as
   \textit{NuSTAR} observations of 2014 outburst and found a truncated accretion disc geometry when the source was in the LHS. Both the results from \citet{Stiele and Yu 2016} and \citet{Ingram et al. 2017} come to an agreement that the source stayed in the LHS during the 2014 outburst.
   
Since two observations at different epochs jointly performed with \textit{XMM-Newton} and \textit{NuSTAR} 
during the 2016 outburst of H~1743--322 are still unexplored, it is worth examining
 in detail the behavior of the source in the light of various characteristics discussed above. Moreover, the 2016 outburst appears to be different from that of the 2008 and 2014 failed outbursts, as the 2016 outburst exhibits a full spectral state transition.
  In this paper, 
 we carry out a systematic spectral and temporal study of H~1743--322 using these observations. Timing study using the \textit{NuSTAR} data 
 allows us to probe the nature of the timing features in the PDS beyond the energy range of the \textit{XMM-Newton}.
 We report the detection of a low-frequency QPO along with  upper harmonic in both epochs. We also find a shift 
 in the centroid frequency of the QPO and the upper harmonic between the two epochs. We have also compared the characteristics of
 the PDS in the high energy band using \textit{NuSTAR} to those obtained from the \textit{XMM-Newton} observations.
Besides, we study the energy dependence of the temporal parameters and compare them with the previous 
 studies. We have also found a hard lag and a log-linear increase of the time-lag with energy in the energy dependent time-lag spectra derived 
 from \textit{XMM-Newton} observations. In addition,
 we present detailed broad-band spectral analysis of joint \textit{XMM-Newton} and \textit{NuSTAR} 
 spectral data in the 2.5--78 keV band, as well as study the accretion disc and the relativistic reflection. We have also studied the relation between spectral and temporal parameters, and discussed
  the possible origin of the variability in the system.

The remainder of the paper is organized as follows. We present the observations and data reduction in section 2, whereas section 3 contains 
the analysis and results of our spectral and temporal study. Finally, section 4 is devoted to discussion and concluding remarks.

\begin{table*}[!ht]
	\caption{\textit{XMM-Newton} and \textit{NuSTAR} observations of H~1743--322 during 2016 outburst.}
   \centering
   \begin{tabular}{lccccc} 
      \hline
      \hline

	Obs. No. & Instrument & Obs. ID. & Obs. date & Exp.(ks) & Obs. mode \\
	\hline
	Epoch - 1 & \textit{XMM-Newton}/EPIC-pn & 0783540301 & 2016 Mar 13 & 142.6 & Timing \\
	& \textit{NuSTAR}/FPMA,FPMB & 80202012002 & 2016 Mar 13 & 65.8 & Imaging\\
	\\
	Epoch - 2 & \textit{XMM-Newton}/EPIC-pn & 0783540301 & 2016 Mar 15 & 142.6 & Timing \\
	& \textit{NuSTAR}/FPMA,FPMB & 80202012004 & 2016 Mar 15 & 65.6 & Imaging \\
	\hline 
    \end{tabular}
\end{table*}

\begin{figure}[ht!]
\plotone{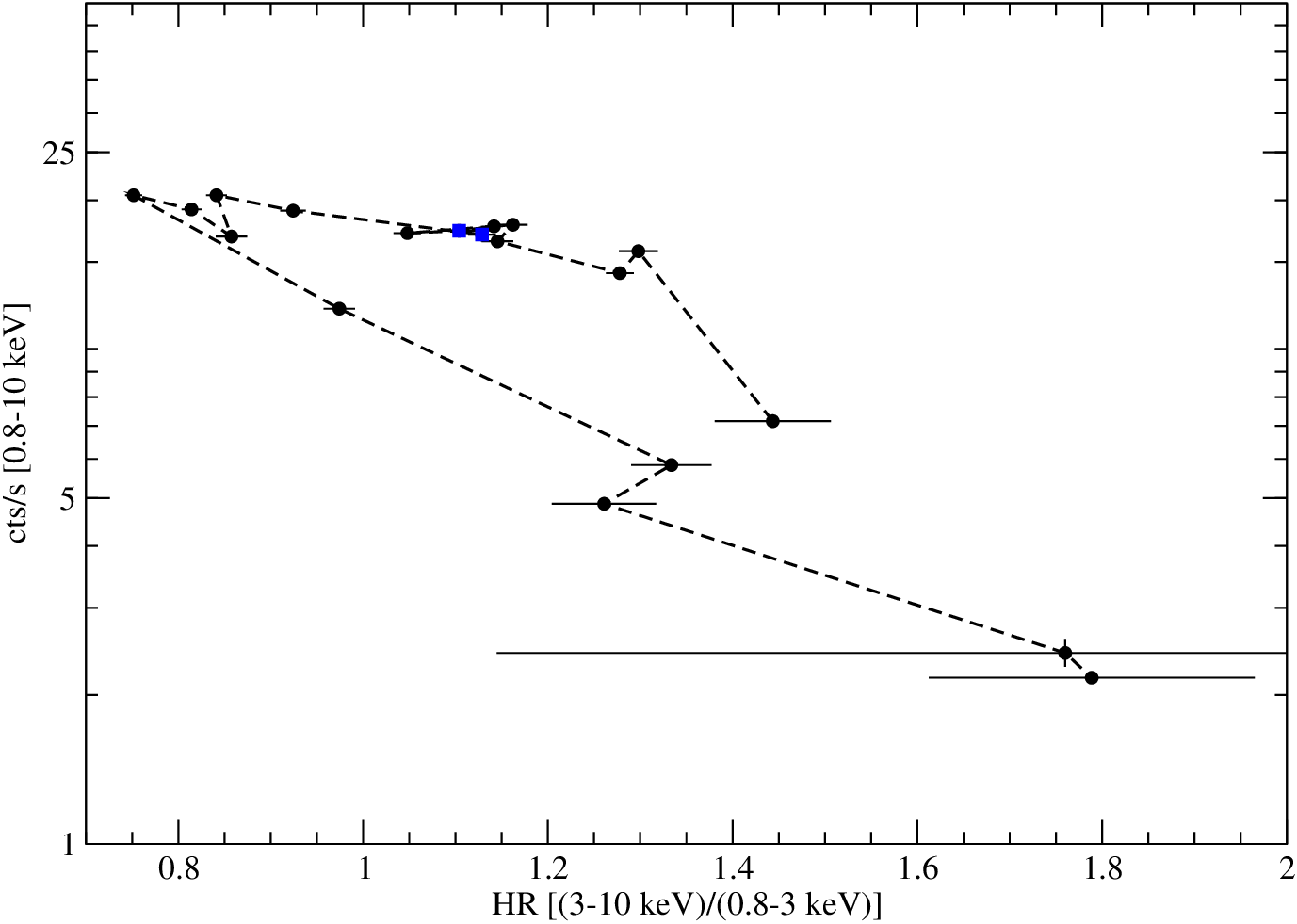}
\caption{Hardness intensity diagram (HID), derived using the window timing-mode data of Swift/XRT from March 01, 2016 to April 08, 2016. The two blue squares indicate the two simultaneous \textit{XMM-Newton} and \textit{NuSTAR} observations on two different epochs, used in this work.}
\end{figure}

\section{OBSERVATIONS AND DATA REDUCTION}

	\subsection{Swift Monitoring}
The 2016 outburst of H~1743--322 was detected and followed by \textit{Swift/XRT} \citep{Burrows2000, Hill2000}. We analyzed all the observations taken in \textit{Swift/XRT} window timing--mode data between March 01, 2016 and April 08, 2016 using the the online data analysis tools provided by the Leicester Swift Data Centre\footnote {\url{http://www.swift.ac.uk/user_objects/}} \citep{Evans2009}. We derived the count rate in the 0.8--10 keV, 0.8--3 keV and 3--10 keV bands, and for the calculation of hardness ratio (hereafter HR), we divided the count rate in the 3--10 keV band by the count rate in the 0.8--3 keV band. Figure 1 shows the hardness intensity diagram (hereafter HID), which depicts that the source undergoes a full state transition during the 2016 outburst.

	\subsection{\textit{XMM-Newton}}
\textit{XMM-Newton} observed  H~1743--322 twice on March 13, 2016 (hereafter Epoch - 1) and March 15, 2016 (hereafter Epoch - 2) for an exposure time of 142.6 ks each. Details of the \textit{XMM-Newton} observations are given in Table 1.
During these two observations, only the European Photon Imaging Camera (EPIC-pn) was employed in the timing mode with a thick filter to observe the source. Scientific Analysis System (SAS v.16.0.0) with the most recent calibration files were utilized to filter and produce the EPIC-pn event files. We did not find any soft proton background flaring from the extracted lightcurve of Epoch - 1 in the energy range of 10--12 keV. However, background flaring towards the end of the observation was detected in the 10--12 keV lightcurve of Epoch - 2. 
To remove the flaring, we created a good time interval (GTI) file with count rate $\leqslant$ $1.7 s^{-1}$. We then applied this GTI file and filtered the event list for the background flaring. We used a rectangular region of 17 pixels width $(29 \leqslant RAWX \leqslant 46)$, keeping the source in the centered position, and extracted the source spectra. A narrow rectangular region of 5 pixels width $(05 \leqslant RAWX \leqslant 10)$ towards the edge of the detector was used to extract the background spectra. Using the \texttt{epatplot} task within SAS, we found that both Epoch - 1 and Epoch - 2 observations were affected by pile-up. In order to reduce the pile-up effect, we used only the single pixel events (\textsc{pattern==0}), as well as also excised the three central columns from the source position of both the observations.  We then generated Redistribution Matrix File (RMF) and Ancilliary Region File (ARF) for each epoch using the tasks \texttt{rmfgen} and \texttt{arfgen} available within SAS, respectively. Moreover, we also extracted the source and background lightcurves  and corrected the source lightcurves for the background contribution using the SAS task '\texttt{epiclccorr}'.

	\subsection{\textit{NuSTAR}}	
	
\textit{NuSTAR} \citep{Harrison et al. 2013} also observed  H~1743--322 simultaneously with \textit{XMM-Newton} in the two epochs and the details of the observations are listed in Table 1. We reduced the data using the \texttt{nupipeline} task available within \textit{NuSTARDAS} with the latest available calibration files. For both the FPMA and FPMB modules, we used a circular region of 30 arcsec by keeping the source in the center and generated the source spectra. We used another circular region of the same size on the same detector away from the source to extract the background. We generated the corresponding RMF and ARF files using the task \texttt{nuproducts} available within \textit{NuSTARDAS}.

\begin{figure}[ht!]
\plotone{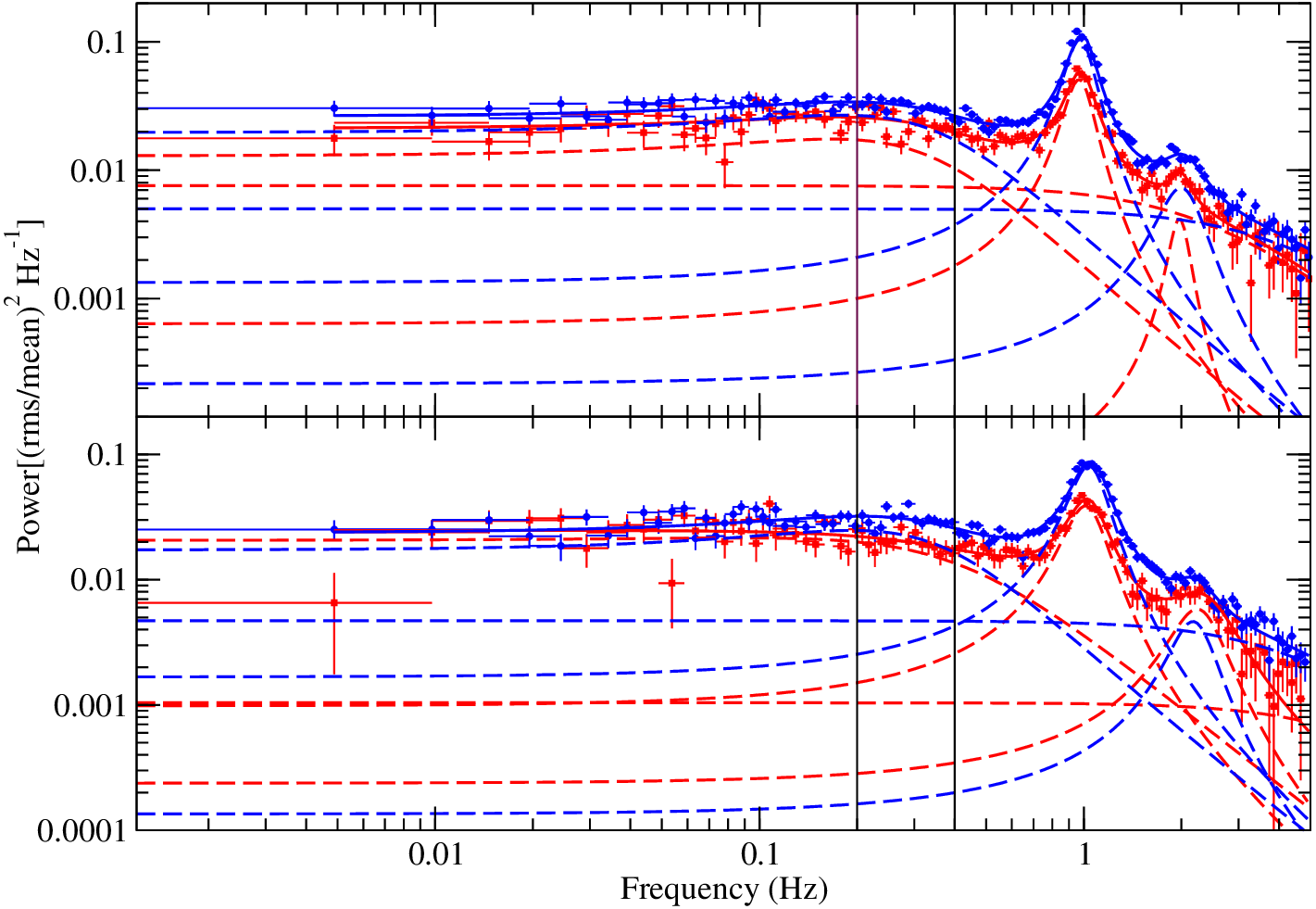}
\caption{Power Density Spectra (PDS) in the 0.7--3 keV and 3--10 keV bands derived from the \textit{XMM-Newton} observations for the Epoch - 1 (upper panel) and Epoch - 2 (lower panel). Data points with filled squares (red) and filled circles (blue) represent the PDSs extracted in the 0.7--3 keV and 3--10 keV bands, respectively. The solid lines indicate the best fitted model. The dashed lines show the individual model components. The two vertical solid lines indicate the frequency range (0.2--0.4 Hz) where the average time-lags have been estimated (see Section 3.2).}
\end{figure}

\begin{figure}[ht!]
\plotone{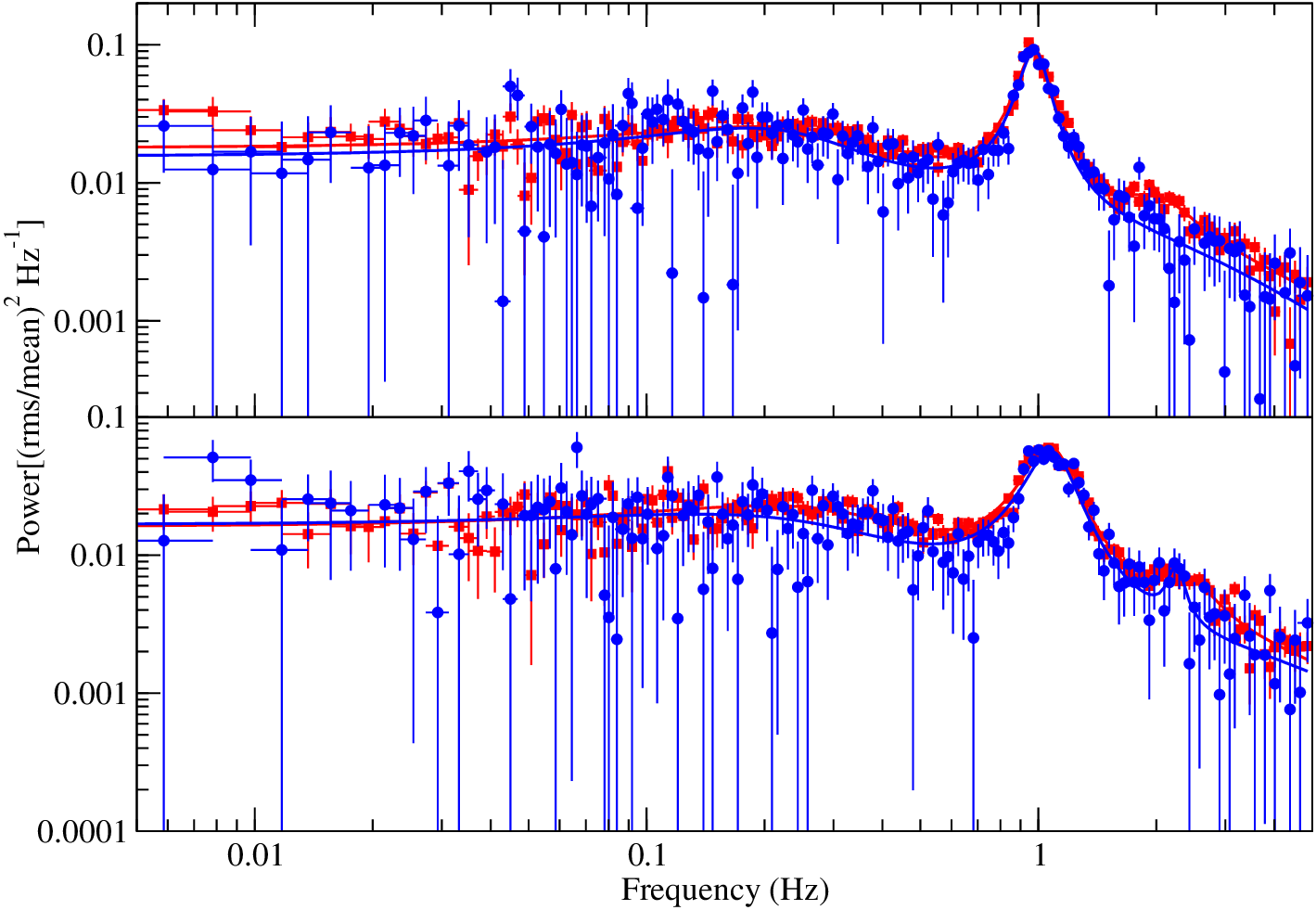}
\caption{Cross Power Density Spectra (CPDS) in the 3--10 keV and 10--30 keV bands derived from the \textit{NuSTAR} observations for the Epoch - 1 (upper panel) and Epoch - 2 (lower panel). Here, data points with filled squares (red) and filled circles (blue) represent the PDSs extracted in the 3--10 keV and hard 10--30 keV bands, respectively. The solid lines indicate the best fitted model.}
\end{figure}

\begin{table*}[!ht]
	\caption{Best-fit temporal parameters obtained from the PDS of the \textit{XMM-Newton} observations.}
   \centering
   \begin{tabular}{lcccc} 
      \hline
      \hline
     & \multicolumn{2}{c}{0.7--3 keV} & \multicolumn{2}{c}{3--10 keV} 
     \\\hline
     & Epoch - 1 & Epoch - 2 & Epoch - 1 & Epoch - 2 \\
     \hline
     $\nu_{qpo}$(Hz) & 0.980$\pm$0.005 & 1.020$\pm$0.009 & 0.980$\pm$0.003 & 1.040$\pm$0.003 \\ 
     $FWHM_{qpo}$(Hz) & 0.22$\pm$0.02 & 0.33$\pm0.03$ & 0.22$\pm$0.01 & 0.30$\pm0.01$ \\ 
     $Q_{qpo}$ & 4.50$^{+0.38}_{-0.44}$ & 3.1$^{+0.2}_{-0.3}$ & 4.4$\pm$0.2 & 3.5$\pm0.1$ \\ 
     $rms_{qpo}$[\%] & 13.0$\pm0.6$ & 13.8$^{+0.5}_{-0.7}$ & 18.8$\pm$0.3 & 19.24$\pm$0.2 \\ 
     $\nu_{har}$(Hz) & 2.0$^{+0.07}_{-0.05}$ & 2.26$\pm0.08$ & 2.01$\pm0.04$ & 2.20$\pm0.05$ \\ 
     $FWHM_{har}$(Hz) & 0.3$^{+0.3}_{-0.2}$ & 0.9$^{+0.4}_{-0.3}$ & 0.7$\pm0.2$ & 0.76 (f) \\ 
     $Q_{har}$ & 6.5$^{+3.0}_{-6.9}$ & 2.5$^{+1.5}_{-1.2}$ & 2.8$^{+0.6}_{-0.8}$ & 2.9$\pm0.1$ \\ 
     $rms_{har}$[\%] & 4.5$^{+0.02}_{-0.01}$ & 8.7$^{+1.9}_{-2.0}$ & 8.7$^{+1.3}_{-1.4}$ & 7.2$^{+0.6}_{-0.7}$ \\ 
     $\nu_{bln}$(Hz) & 0.17$^{+0.03}_{-0.04}$ & 0.09$\pm0.05$ & 0.18$\pm0.02$ & 0.19$\pm0.01$ \\ 
     $FWHM_{bln}$(Hz) & 0.59$\pm0.04$ & 0.8 (f) & 0.59$^{+0.08}_{-0.07}$ & 0.59$^{+0.06}_{-0.05}$ \\ 
     $Q_{bln}$ & 0.29$\pm$0.03 & 0.11$\pm0.02$ & 0.300$^{+0.010}_{-0.004}$ & 0.330$^{+0.007}_{-0.001}$ \\ 
     $rms_{bln}$[\%] & 10.3$^{+1.8}_{-2.2}$ & 12.4$^{+0.9}_{-1.8}$ & 13.0$^{+0.7}_{-1.0}$ & 12.5$^{+0.4}_{-0.5}$\\ 
	 $\nu_{bln_{zero}}$(Hz) & 0 (f) & 0 (f) & 0 (f) & 0 (f) \\ 
	 $FWHM_{bln_{zero}}$ & 4.9$^{+3.2}_{-1.4}$ & $10^p$ & 8.5$^{p}_{-2.5}$ & 9.2$^{p}_{-1.9}$  \\ 
	 $rms_{bln_{zero}}$[\%] & 17.0$^{+1.5}_{-1.7}$ &  10.5$^{+3.2}_{-3.9}$ & 18.3$^{+1.1}_{-1.2}$ &  18.4$\pm0.9$\\    
     \hline
   \end{tabular}
   \begin{tablenotes}
		\item f -- indicates the fixed parameters, p -- indicates the parameters pegged at lower/upper bounds
		\end{tablenotes}
\end{table*}

\begin{table*}[!ht]
	\caption{Best-fit temporal parameters obtained from the PDS of the \textit{NuSTAR} observations.}
   \centering
   \begin{tabular}{lcccc} 
      \hline
      \hline
     & \multicolumn{2}{c}{3--10 keV} & \multicolumn{2}{c}{10--30 keV} 
     \\\hline
     & Epoch - 1 & Epoch - 2 & Epoch - 1 & Epoch - 2 \\
     \hline
	$\nu_{qpo}$(Hz) & 0.980$\pm$0.003 & 1.060$\pm$0.005 & 0.980$\pm$0.005 & 1.07$\pm$0.01 \\ 	
	$FWHM_{qpo}$(Hz) & 0.20$\pm$0.01 & 0.35$\pm0.02$ & 0.18$\pm$0.02 & 0.34$\pm0.03$ \\ 
	$Q_{qpo}$ & 4.9$^{+0.25}_{-0.27}$ & 3.1$\pm0.12$ & 5.4$^{+0.4}_{-0.5}$ & 3.1$^{+0.2}_{-0.3}$ \\ 
	$rms_{qpo}$[\%] & 16.1$\pm0.4$ & 17.1$\pm0.3$ & 15.4$\pm$0.6 & 16.5$^{+0.2}_{-0.6}$ \\ 	
	$\nu_{har}$(Hz) & 2.01$\pm0.1$ & 2.37$\pm0.01$ & ... & 2.25$^{+0.07}_{-0.80}$ \\ 
	$FWHM_{har}$(Hz) & 1.02 (f) & 0.99 (f) & ... & 0.23 (f) \\ 
	$Q_{har}$ & 1.96$\pm0.1$ & 2.4$\pm0.1$ & ... & 9.7$\pm0.3$ \\	
	$rms_{har}$[\%] & 7.13$\pm1.1$ & 7.2$^{+0.7}_{-0.8}$ & ... & 4.4$^{+1.0}_{-1.3}$ \\	
	$\nu_{bln}$(Hz) & 0.19$\pm0.02$ & 0.19$\pm0.02$ & 0.18$\pm0.03$ & 0.13 (f) \\ 
	$FWHM_{bln}$(Hz) & 0.5$\pm0.01$ & 0.5$\pm0.1$ & 0.35$^{+0.13}_{-0.20}$ & 0.54$^{+0.16}_{-0.12}$ \\ 
	$Q_{bln}$ & 0.38$^{+0.3}_{-0.2}$ & 0.360$^{+0.100}_{-0.001}$ & 0.52$^{+0.08}_{-0.06}$ & 0.23$^{+0.05}_{-0.07}$ \\	
	$rms_{bln}$[\%] & 10.3$^{+0.9}_{-2.2}$ & 12.4$^{+0.9}_{-1.0}$ & 8.5$^{+1.2}_{-1.5}$ & 9.2$^{+0.9}_{-1.0}$ \\	
	$\nu_{bln_{zero}}$(Hz) & 0 (f) & 0 (f) & 0 (f) & 0 (f) \\	
	$FWHM_{bln_{zero}}$ & 6.6$^{p}_{-1.72}$ & $10^p$ & 4.5$^{+2.6}_{-1.3}$ & 10$^p$  \\
	$rms_{bln_{zero}}$[\%] & 15.5$^{+1.1}_{-1.2}$ &  14.8$^{+1.3}_{-1.5}$ & 15.0$^{+1.4}_{-1.6}$ &  14.0$^{+2.5}_{-3.1}$ \\ 
	\hline
		\end{tabular}
		\begin{tablenotes}
		\item f -- indicates the fixed parameters, p -- indicates the parameters pegged at lower/upper bounds
		\end{tablenotes}
\end{table*}

\section{ANALYSIS AND RESULTS}

\subsection{Power Density Spectra}

For the timing analysis, we used the Interactive Spectral Interpretation System (ISIS, V.1.6.2--40) \citep{HouckandDenicola2000}. We quote the errors at 90\% confidence level.
The PDSs from the background subtracted lightcurves of the \textit{XMM-Newton}/EPIC-pn observations were computed using ``POWSPEC" task within FTOOLS in the two energy bands of  0.7--3 keV and 3--10 keV. After subtracting the contribution due to Poisson noise \citep{Zhangetal1995}, we normalized the PDSs according to \citet{Leahyetal1983} and then converted the variability power to the square fractional rms \citep{Belloni and Hasinger 1990}.
Figure 2 shows the PDSs in the 0.7--3 keV and 3--10 keV bands for Epoch - 1 and Epoch - 2, where the presence of the low-frequency QPO along with upper harmonic at $\sim$1Hz and $\sim$2Hz are prominent. The PDSs in both the energy bands exhibit identical shape and require a zero centered Lorentzian, as well as three peaked Lorentzian components for the QPO, upper harmonic and the band limited noise (BLN) component. All the best-fit parameters are listed in Table 2. 
The significance of detection of the QPO (upper harmonic) in the 0.7--3 keV band is 17.9$\sigma$ (3.1$\sigma$) and 19$\sigma$ (4.1$\sigma$), whereas that found in the 3--10 keV  band is 43.4$\sigma$ (5.2$\sigma$) and 40.7$\sigma$ (6.6$\sigma$) for Epoch - 1 and Epoch - 2, respectively. This suggests that the significance of both the QPO and the upper harmonic remains more or less similar between Epoch - 1 and Epoch - 2 for each energy band, whereas their values are found to be increasing with energy band in each epoch. 
From Table 2 it is clear that in both the observations, the QPO and the upper harmonic are detected at $\sim$1:2 ratio in each energy band. We also note that the centroid frequencies of the QPO and the upper harmonic do not change with energy in both the epochs.
However, we found that the centroid frequencies of the QPO (upper harmonic) in Epoch - 2 are shifted towards higher frequencies by $0.04\pm0.01$ Hz ($0.26\pm0.1$ Hz) in the 0.7--3 keV band and $0.06\pm0.004$ Hz ($0.19\pm0.06$ Hz) in the 3--10 keV band with respect to those found in Epoch - 1.
The quality factor ($Q$=$\nu_{centroid}$/FWHM; FWHM - full width at half maximum) of the QPO in each energy band is reduced in Epoch - 2 compared to that in Epoch - 1, however, the 
$Q$--factor for the upper harmonic remains almost similar for both the energy bands and epochs.
Although the fractional rms variability of the QPO does not change between Epoch - 1 and Epoch - 2 within each energy band, it is found to be 
higher in the 3--10 keV band with respect to that obtained in the 0.7--3 keV band for each epoch (see Table 2).
Apart from this, the fractional rms variability of the upper harmonic appears to be larger in Epoch - 2 than that in Epoch - 1 in the 0.7--3 keV band, whereas it remains the same for both the epochs in 3--10 keV band.

In addition to the above, we have also carried out the timing analysis of H~1743--322 using the \textit{NuSTAR} observations. For this analysis, we have considered only the 3--30 keV band as this band encompasses 97\% of the source photons detected by \textit{NuSTAR} in the 3--78 keV energy range \citep{Stiele and Kong 2017}. We derived the cross-power density spectra (CPDS) in the 3--10 and 10--30 keV energy bands using MaLTPyNT \citep{Bachetti2015}. The signals from the two completely independent focal plane modules are used to generate the CPDS, which acts like a good alternative of the white noise subtracted PDS \citep{Bachetti2015}.
For the generation of the CPDS, we used the time bins of 0.1 s with the stretches of 512 s. Figure 3 shows the CPDSs derived from the 
\textit{NuSTAR} observations for the Epoch - 1 and Epoch - 2 in the 3--10 and 10--30 keV bands. The best-fit parameters are given in Table 3. 
The shape of the CPDS and the required best-fit model for both the epochs in the 3--10 keV  band are found to be the same as obtained in the same energy band of \textit{XMM-Newton} observations. The ratio at which the QPO and the upper harmonic is detected in each energy band are found to be consistent with the \textit{XMM-Newton} observations. The shifts in the centroid frequencies of the QPO and the upper harmonic in Epoch - 2 with respect to those in Epoch - 1 are $0.08\pm0.006$ Hz and $0.36\pm0.1$ Hz towards the higher frequency side. As can be seen from the Table 2 and 3 that the value of the quality factor ($Q$) and the fractional rms variability of the QPO, as well as the upper harmonic in both the epochs show similar nature to those obtained from the \textit{XMM-Newton} observations in the same energy band. Contrary to the \textit{XMM-Newton} observations in this energy band, the significance level of the QPO, as well as the upper harmonic 
increase from 36.3 $\sigma$ to 47 $\sigma$ for the QPO and 5.4 $\sigma$ to 8 $\sigma$ for the upper harmonic in Epoch - 2 with respect to Epoch - 1.

The \textit{NuSTAR} CPDS in the 10--30 keV band shows the similar shape as in the 3--10 keV band. The QPO and upper harmonic in this band are found in the $\sim1:2$ ratio in Epoch - 2 similar to the CPDS in the 3--10 keV band. However, no signature of the upper harmonic was found in the 10--30 keV band of Epoch - 1 that may be due to lower S/N of CPDS compared to that in Epoch - 2. The shift in the centroid frequency of the QPO in Epoch - 2 with respect to Epoch - 1 resembles to that obtained in the 3--10 keV band of the \textit{NuSTAR} observations and is $0.09\pm0.01$ Hz towards the higher frequency side. The fractional rms amplitude of the QPO is similar in both the epochs in the 10--30 keV band and is consistent with those found in the 3--10 keV band. On the other hand, the upper harmonic in the Epoch - 2 exhibits a reduced fractional rms amplitude in the 10--30 keV band compared to the 3--10 keV band (see Table 3). The $Q$--factor of the QPO also exhibits similar behavior as  the 3--10 keV energy band.

\begin{figure}[ht!]
\plotone{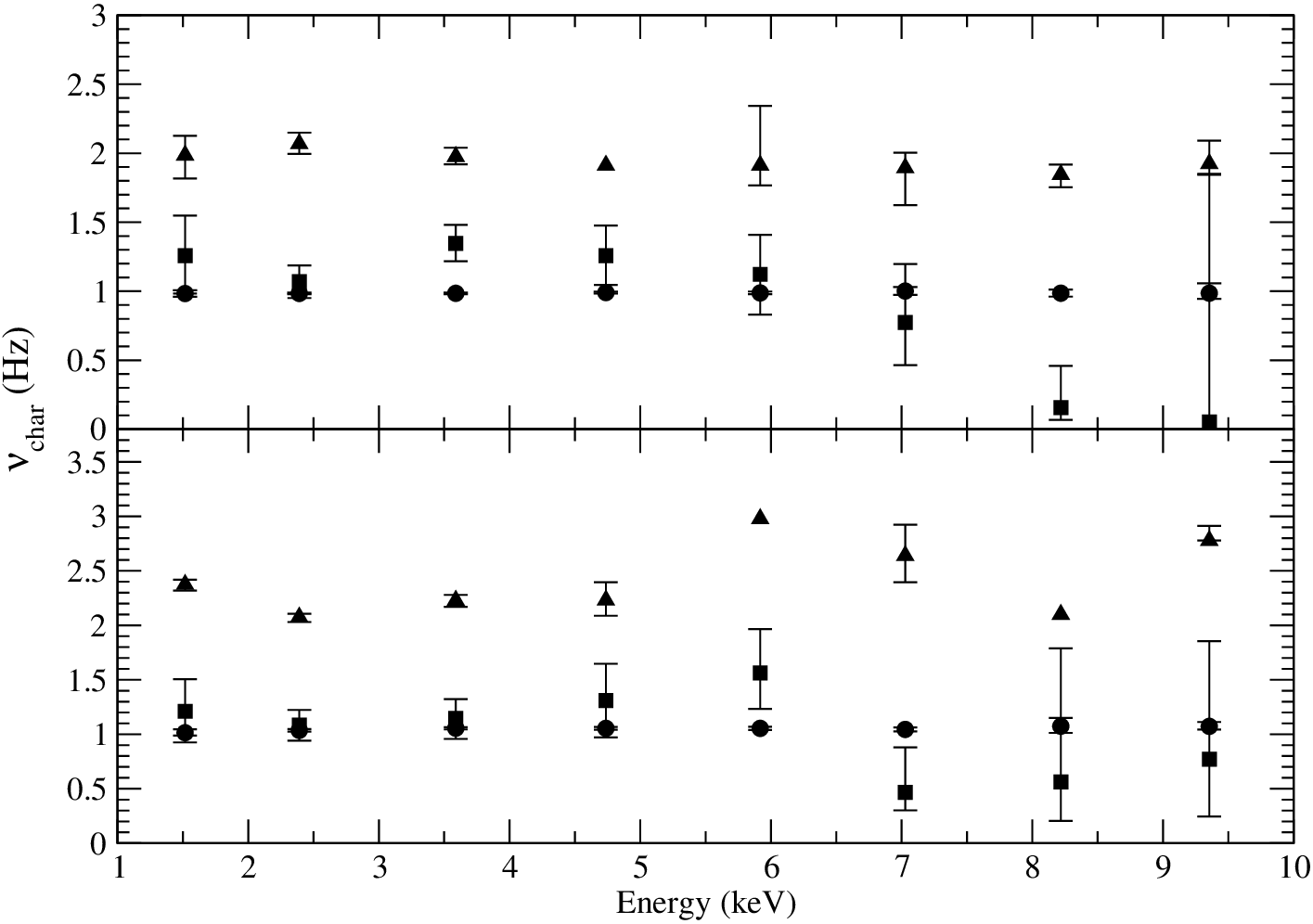}
\caption{Evolution of the Characteristics frequencies of QPO (filled circles), harmonic (filled triangles) and zero centered BLN (filled squares) with energy for the Epoch - 1 (upper panel) and Epoch - 2 (lower panel).}
\end{figure}

\begin{figure}[ht!]
\plotone{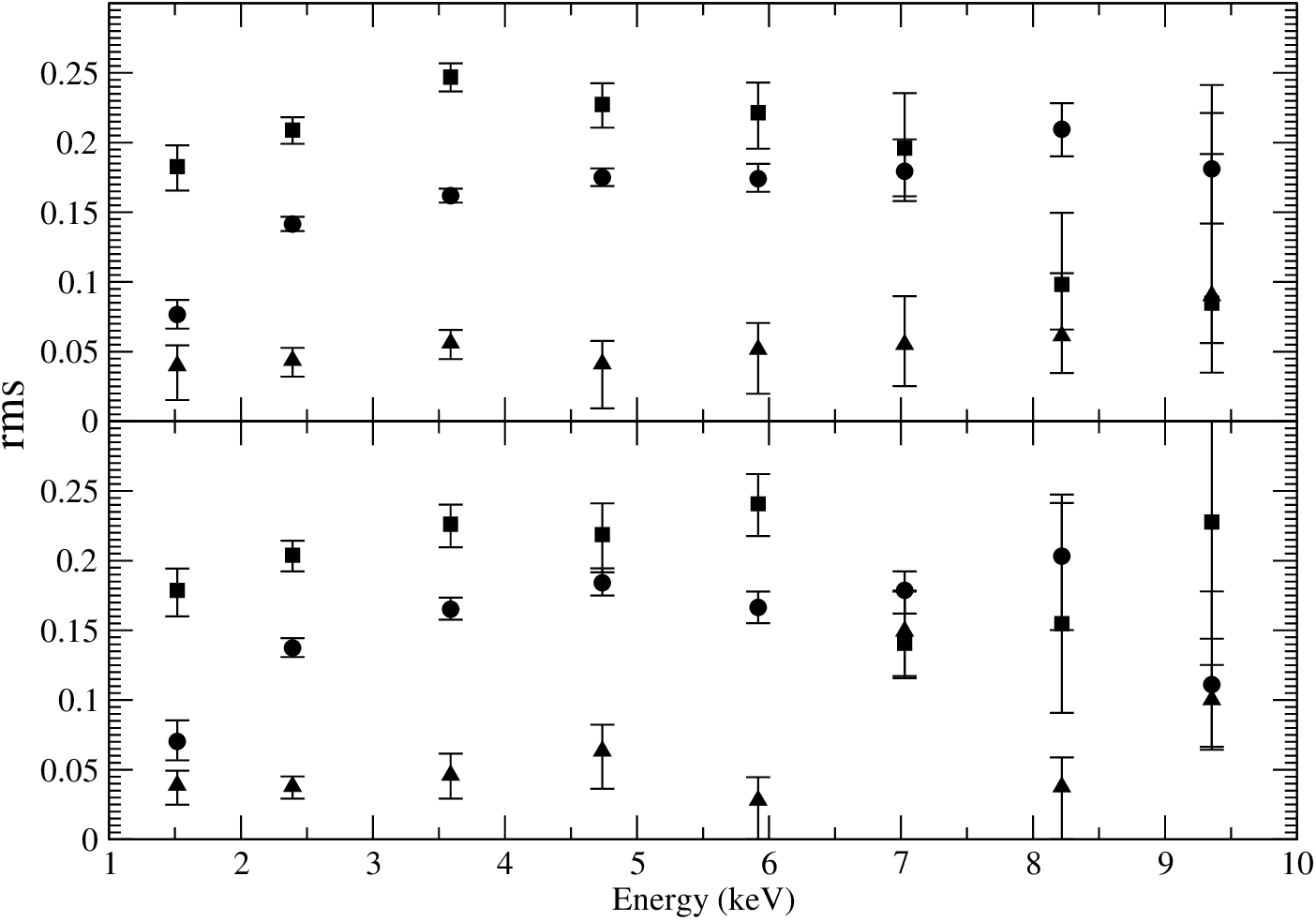}
\caption{Evolution of the fractional rms amplitude of QPO (filled circles), harmonic (filled triangles) and zero centered BLN (filled squares) with energy for the Epoch - 1 (upper panel) and Epoch - 2 (lower panel).}
\end{figure}

In order to study the evolution of characteristic frequency and fractional rms amplitude with energy, we divided the full energy band of the 
\textit{XMM-Newton} observations into 8 equal narrow bands of $\sim$ 1 keV width, and derived Poisson noise subtracted, as well as rms normalized PDS in each energy band. For this analysis, we excluded the \textit{NuSTAR} observations due to low signal to noise ratio in each of these energy bands.
The PDSs derived for each narrow energy band of \textit{XMM-Newton} data were modeled with three Lorentzian components for the QPO, the upper harmonic and the zero centered BLN.
We calculated the characteristics frequency ($\nu_{char}=\sqrt{\nu^2+\Delta^2}$; where $\Delta$ is the half width at half maximum) and fractional rms amplitude of the QPO, upper harmonic and the zero-centered BLN in each energy band for Epoch - 1  and Epoch - 2. Figure 4 shows the evolution of the characteristics frequency of the QPO, as well as its  upper harmonic and the zero centered BLN as a function of energy. It is clear that the characteristics frequencies of the QPO and its upper harmonic
  show a flat nature without showing any significant dependence on energy for both the epochs. The characteristics frequency of the zero centered BLN also remains almost flat except for the slight decrease above $\sim$6 keV. Figure 5 exhibits the rms spectra of the QPO, as well as its upper harmonic and the zero centered BLN component for both the epochs, which demonstrate either flat or slightly decreasing trend with the energy.
 
\subsection{Frequency-dependent Lag and Lag-energy Spectra}

We used only \textit{XMM-Newton} data and the GHATS package\footnote {\url{http://astrosat.iucaa.in/~astrosat/GHATS_Package/Home.html}} for the lag analysis. 
For the study of evaluation of time lag as a function of temporal frequency, we extracted EPIC-pn lightcurves in 1-1.5 keV and 1.5-4 keV bands. Each of these lightcurves was divided into 141 segments each with a length of 983 s. We computed Fourier transform for each segment and calculated average cross-spectrum. Using the averaged cross-spectrum, we calculated the frequency dependent time lag for both the epochs \citep{Uttley2014}. As in the top panels of Figure 6, we found a hard lag of $0.40\pm0.15$ and $0.32\pm0.07$ s between the above mentioned energy bands in the 0.07--0.4 Hz frequency range for Epoch - 1 and Epoch - 2, respectively. It is noteworthy that error bars are dominating below 0.1 Hz, and no time lag has been observed above 0.4 Hz in the time-lag-frequency spectra for both the epochs. The coherence in the 0.2--0.4 Hz frequency range is also found to be closer to unity for both the epochs.
To study the variation of the time lag as a function of energy, we generated lightcurves in the  0.3--0.7, 0.7--1, 1--1.5, 4--5, 5--6, 6--7, 7--8 and 8--10 keV  bands for both the epochs. We considered the 1.5--4 keV band as the reference band. The energy dependent time lag was estimated between the each narrow energy band and the reference band. Figure 6 (bottom panels) depicts the averaged time lags, estimated in the frequency range of 0.2--0.4 Hz, as a function of energy. This exhibits the increasing nature of the time lag with energy in a log-linear trend.

\begin{figure*}
\gridline{\fig{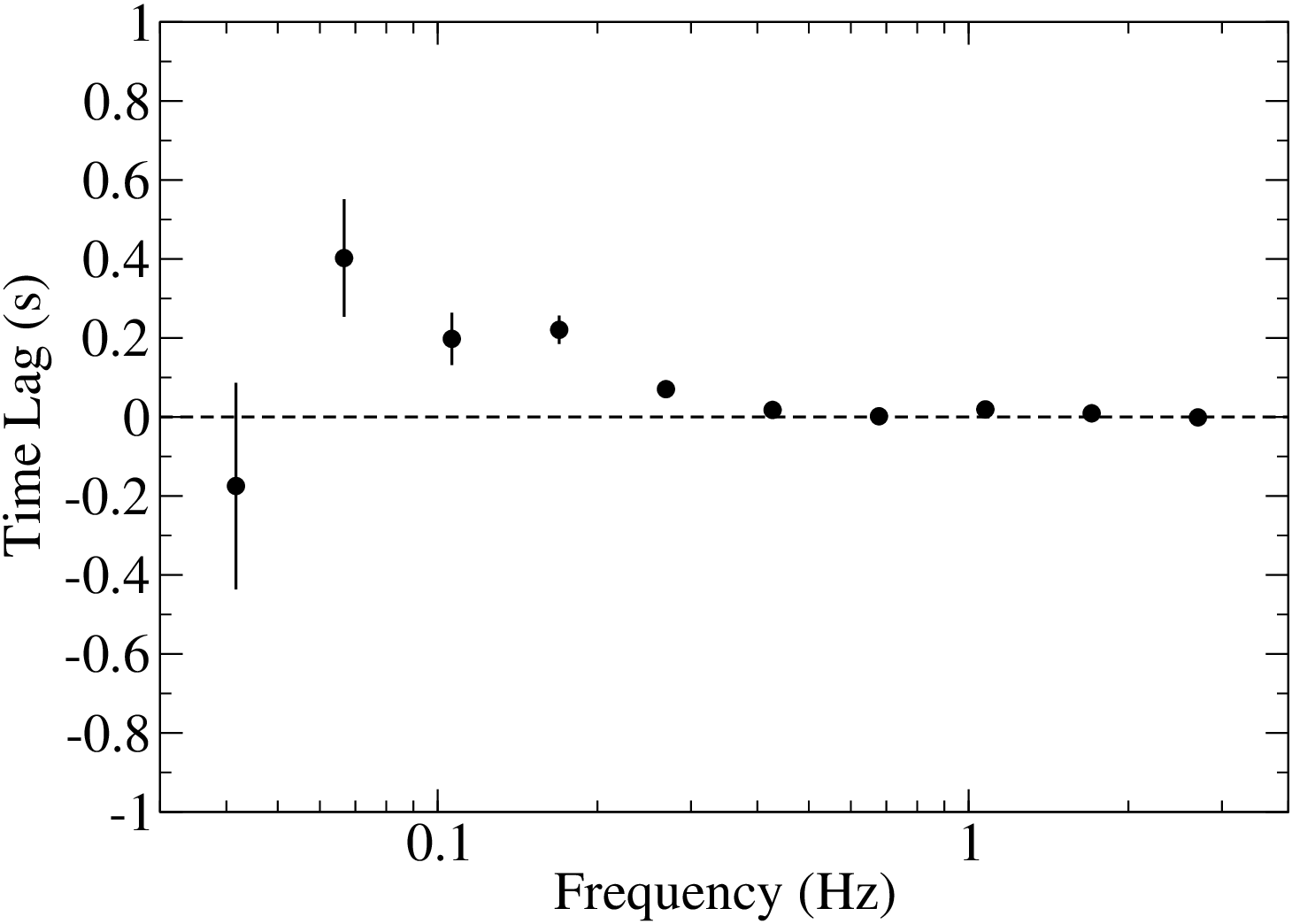}{0.4\textwidth}{}
          \fig{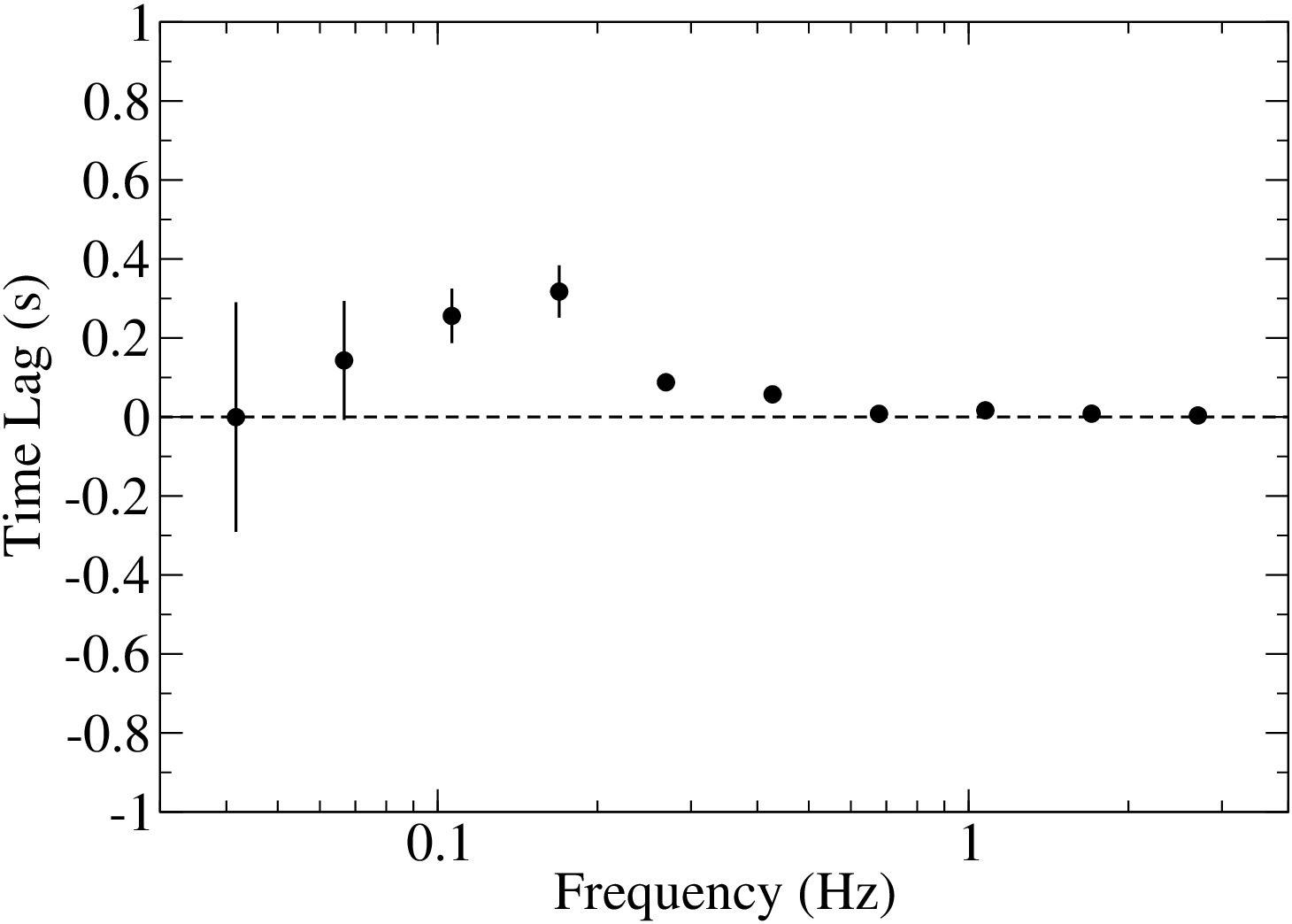}{0.4\textwidth}{}
          }
\gridline{\fig{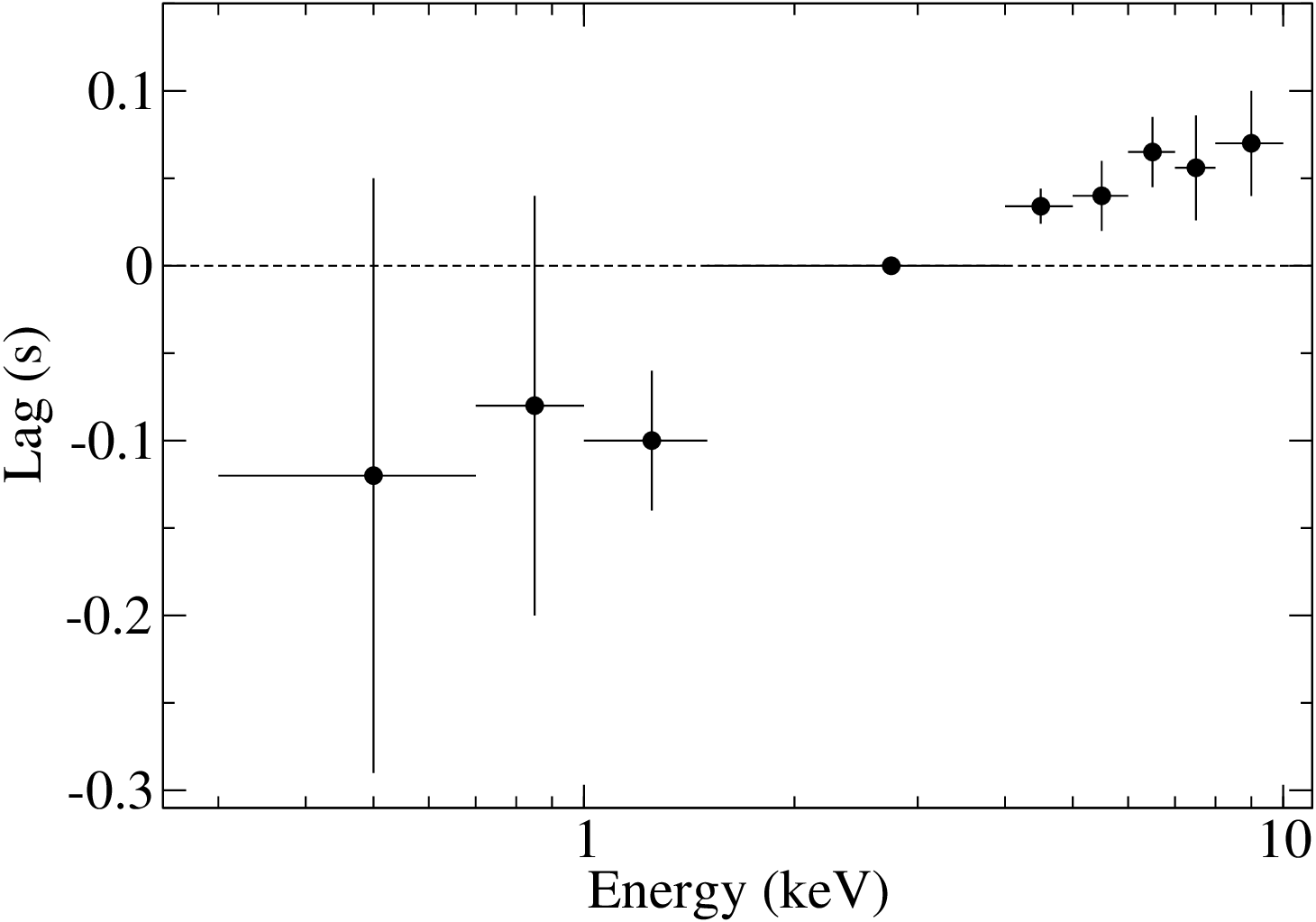}{0.4\textwidth}{}
          \fig{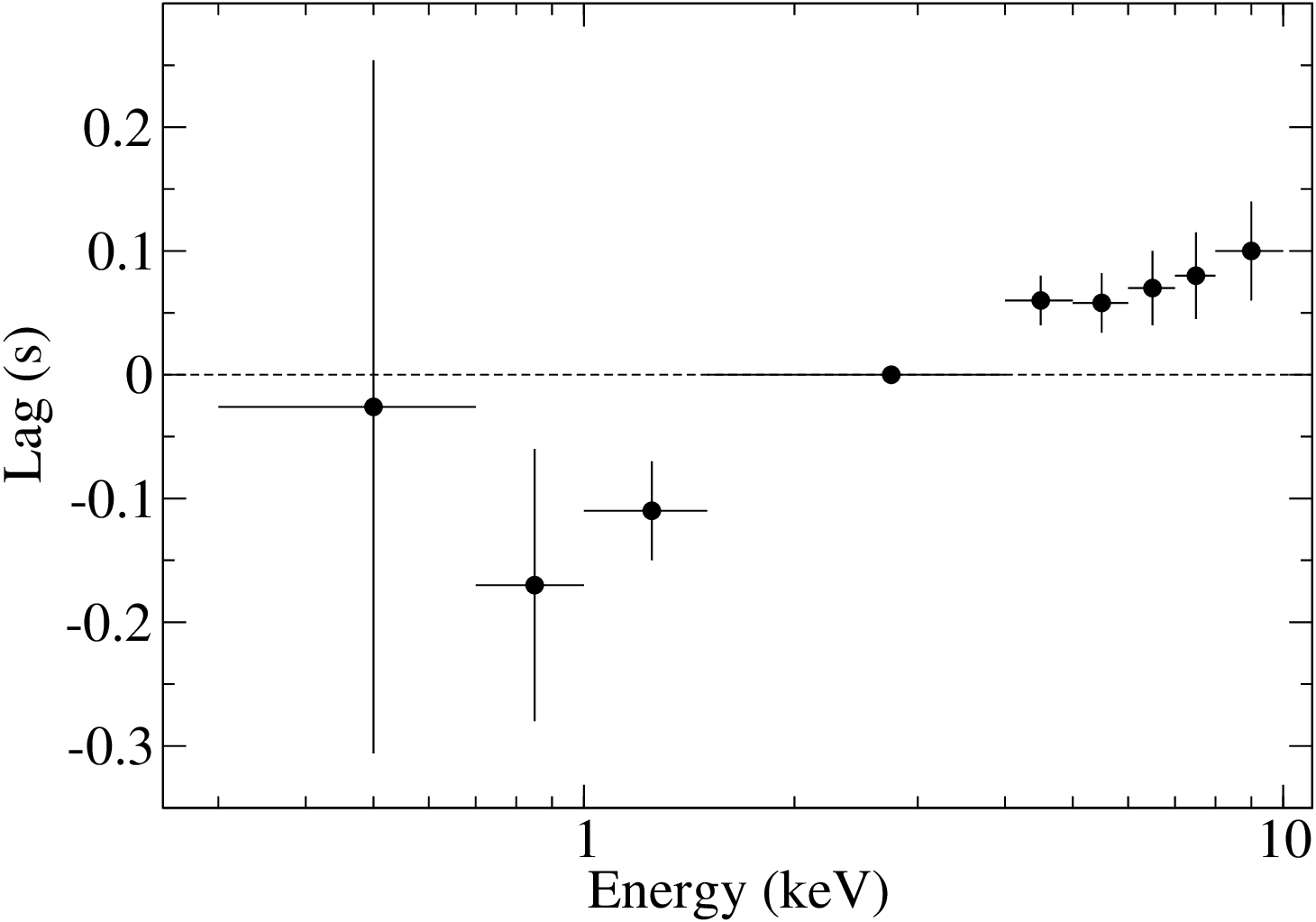}{0.4\textwidth}{}
          }
\caption{Frequency-dependent time lags between 1--1.5 keV and 1.5--4 keV band lightcurves for Epoch - 1 (upper left panel) and Epoch - 2 (upper right panel). Positive lag implies the hard lag. Lag - energy spectra for the averaged lag in the 0.2--0.4 Hz frequency range for the Epoch - 1 (lower left panel) and Epoch - 2 (lower right panel). A log-linear trend of lag with energy can be seen. The reference band is always 1.5--4 keV band corresponding to zero time lag point.}
\end{figure*}
 
 \begin{figure}
\plottwo{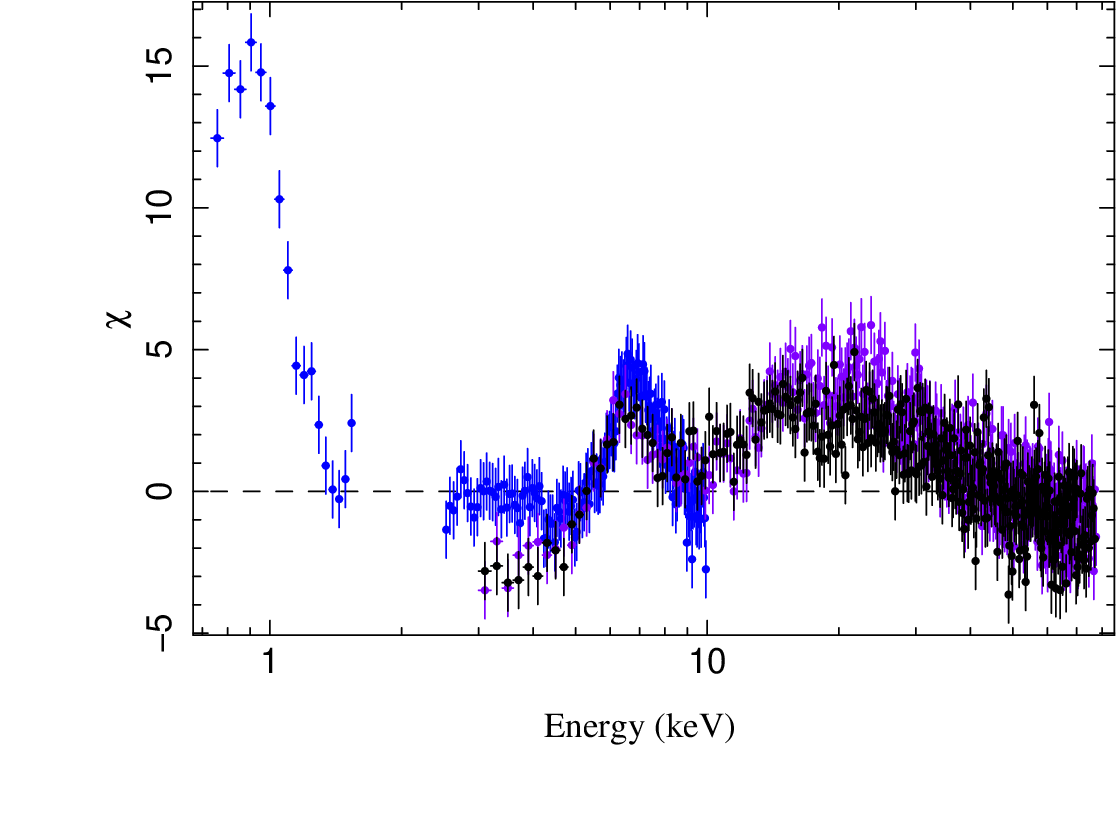}{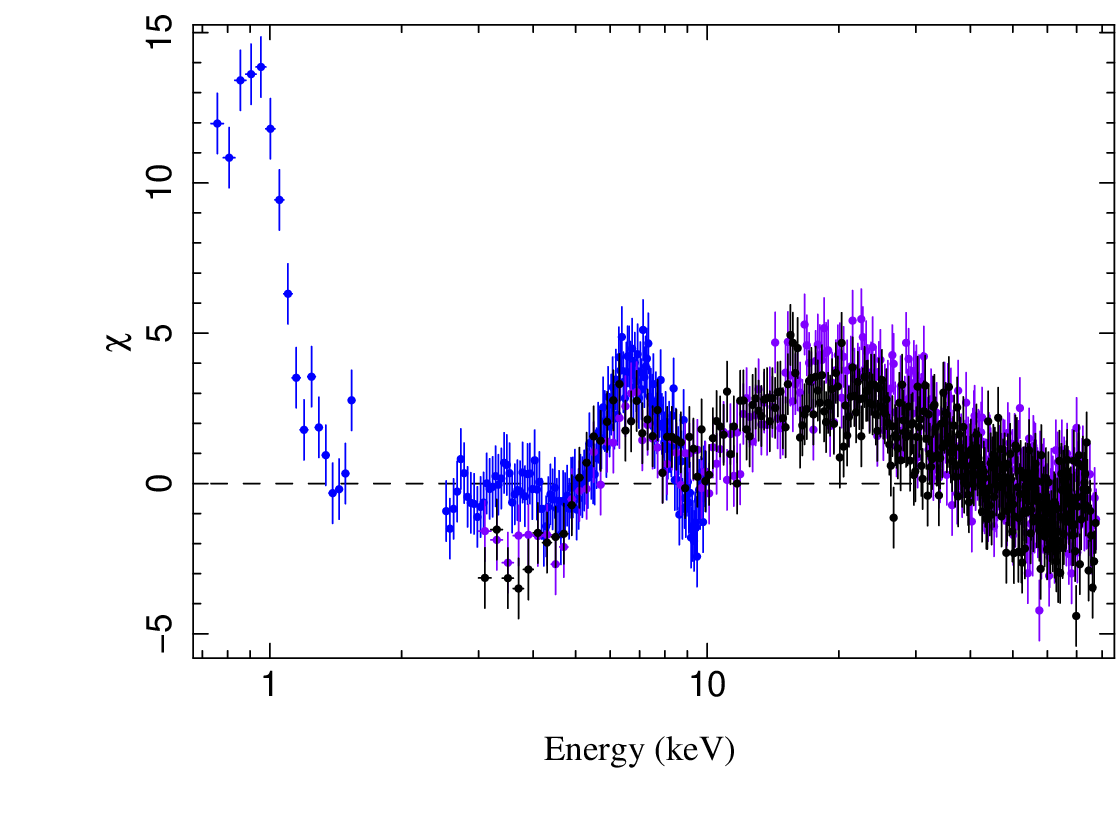}
\caption{The residuals show the deviations of the observed spectral data from the best-fitting CONSTANT*$E^{\Delta{\Gamma}}$*TBABS*(POWERLAW+DISKBB) models. The iron line excess in the 6--8 keV, as well as the reflection hump in the 15--30 keV region are prominent. Blue circles are for \textit{XMM-Newton}/EPIC-pn spectral data, whereas purple and black are for \textit{NuSTAR} FPMA and FPMB spectral data, respectively. The left panel indicates the  Epoch - 1, whereas the right panel represents the Epoch - 2.\label{fig:f2}}
\end{figure} 
  
  \begin{figure}
\plottwo{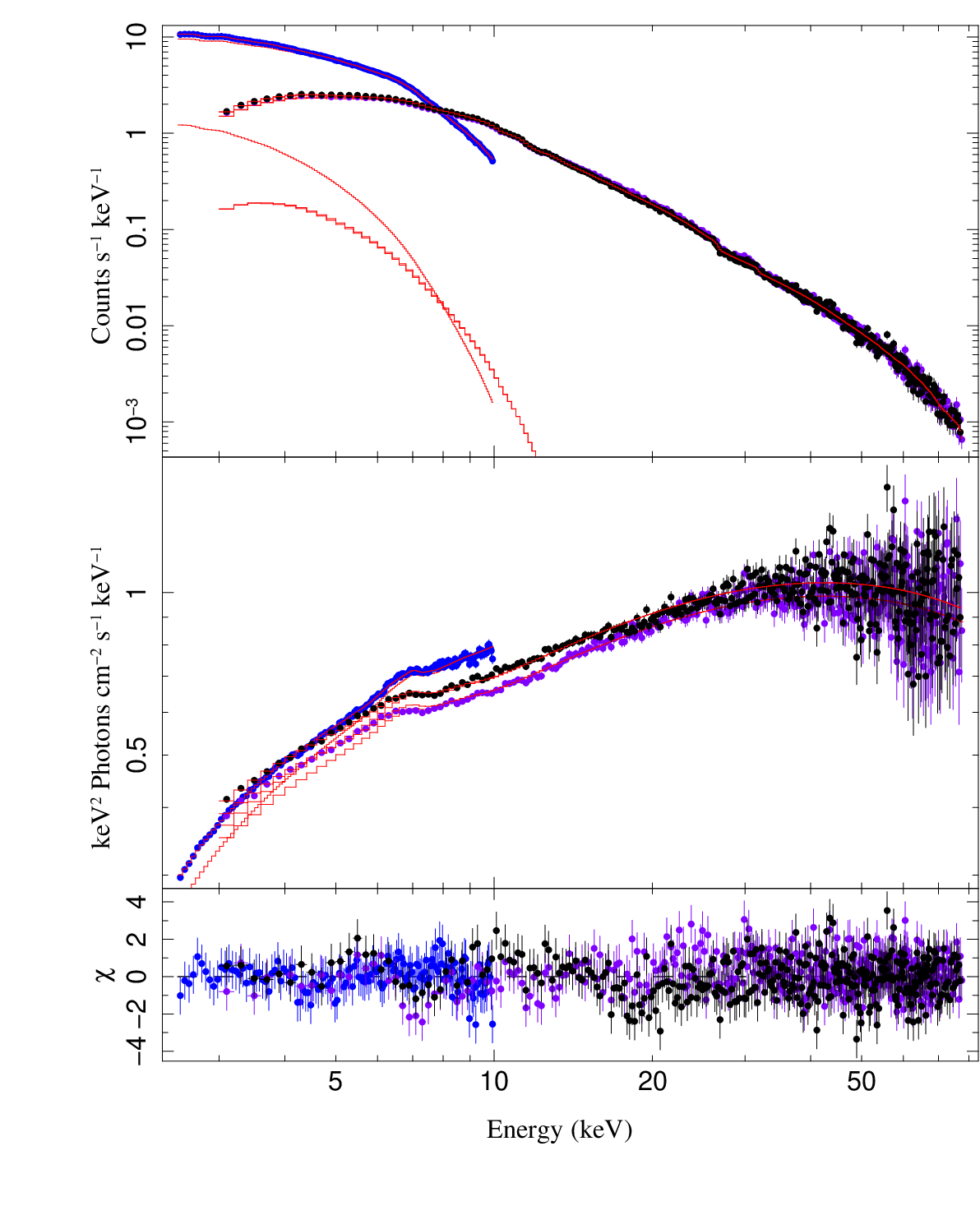}{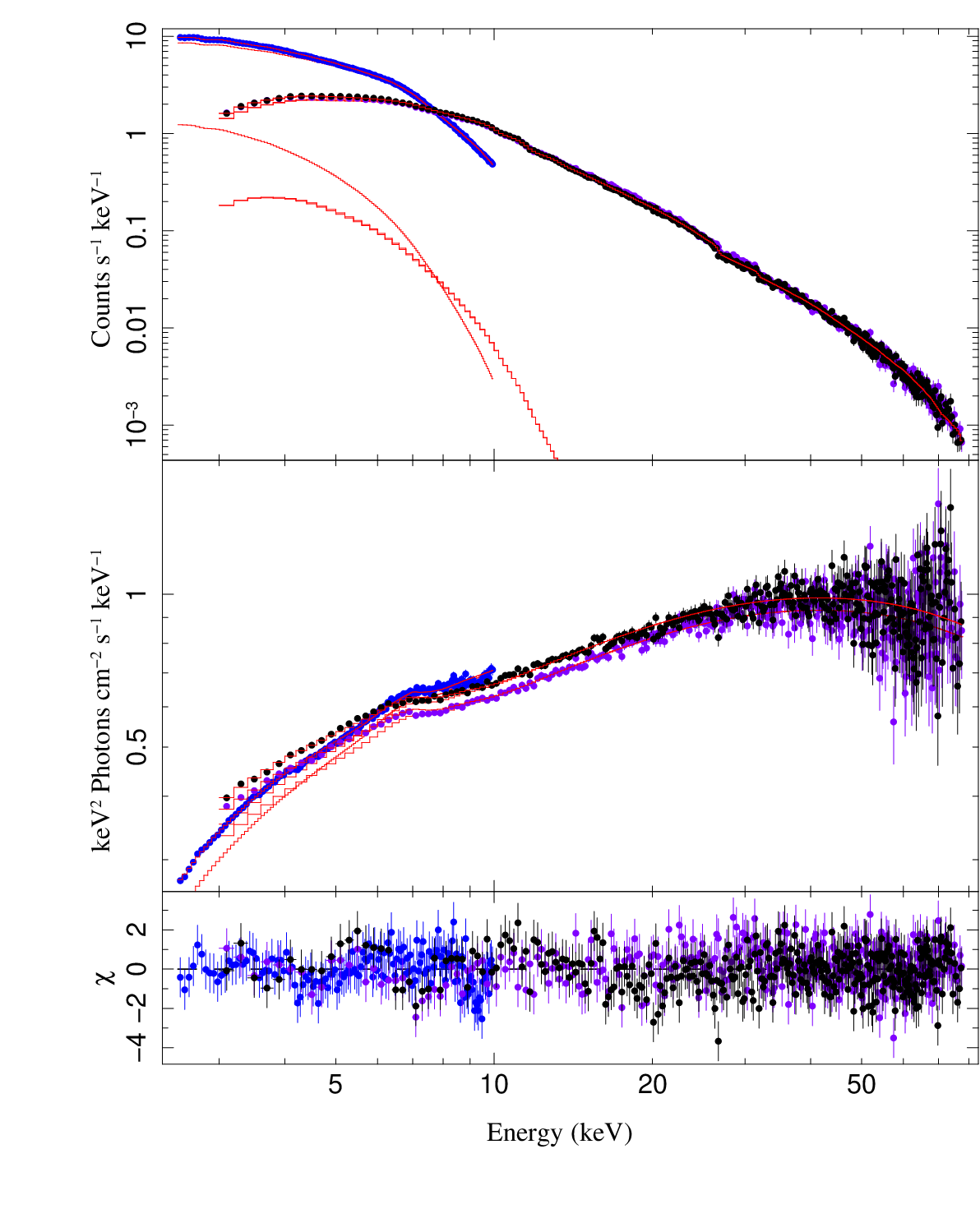}
\caption{Joint \textit{XMM-Newton}/EPIC-pn and \textit{NuSTAR}/FPMA, FPMB counts spectrum (top panel) and unfolded X-ray continuum (middle panel) with the best fitting model CONSTANT*$E^{\Delta{\Gamma}}$*TBABS*(DISKBB+RELXILL) and residual (bottom panel). Blue circles are for the \textit{XMM-Newton}/EPIC-pn spectral data, whereas purple and black are for \textit{NuSTAR} FPMA and FPMB spectral data, respectively. The left panel indicates the Epoch - 1, whereas the right panel represents the Epoch - 2.}
\end{figure}  
 
 \subsection{Energy Spectra}

The time averaged \textit{XMM-Newton}/EPIC-pn spectral data in the 0.7--10 keV band and \textit{NuSTAR} FPMA, FPMB spectral data in the 3--78 keV were fitted simultaneously using ISIS (Version 1.6.2-40). The errors on the best fitted parameters are calculated at 90\% confidence level unless otherwise specified. A systematic uncertainty of 1$\%$ was added to each \textit{XMM-Newton}/EPIC-pn and \textit{NuSTAR} FPMA/FPMB dataset to account for calibration uncertainty between different instruments \citep{Ingram et al. 2017, Madsen2017}.
In order to use $\chi^2$ minimization to obtain the best fit, we grouped the EPIC-pn data to a minimum signal-to-noise ratio of 5 and a minimum of 10 channels per bin. Similarly, we also grouped the FPMA and FPMB data to the same signal-to-noise ratio used for the EPIC-pn data but with a minimum number of channels of 5. Initially, we fitted the three spectral datasets jointly with a POWERLAW  model modified by the Galactic absorption.  We used the absorption model TBabs with the abundances given by \citet{Wilms et al. 2000} and the cross section as in \citet{Verner et al. 1996}.
We also multiplied the absorbed POWERLAW model with a constant factor to account for any difference in the relative normalizations of the three instruments. We fixed the constant factor to 1 for the FPMA data and varied for the EPIC-pn and FPMB data.
We noticed discrepancy between the \textit{XMM-Newton} and \textit{NuSTAR} spectral dataset in the 3--10 keV band for both the epochs. Similar discrepancy was found in the 2014 outburst of H~1743--322 and was eliminated with the inclusion of an additional  $E^{\Delta{\Gamma}}$ model by \citet{Ingram et al. 2017}. We adopted the same procedure in our analysis. We fixed the value of ${\Delta{\Gamma}}$ at zero for both the \textit{NuSTAR} FPMA and FPMB datasets and varied for the EPIC-pn data.
This model provided an unacceptable fit with $\chi^2$/dof equal to $8109.6/884$ and $7699.6/883$ for the Epoch - 1 and Epoch - 2, respectively.
Inclusion of the multicoloured disc blackbody model (DISKBB) \citep{Mitsuda et al. 1984} to 
account for the thermal emission from the accretion disc improved the fit with
$\chi^2$/dof = $6456.8/882$ for Epoch - 1 and $\chi^2$/dof = $6079.6/881$ for the Epoch - 2. However, as shown in Figure 7, the model  CONSTANT*$E^{\Delta{\Gamma}}$*TBABS*(DISKBB+POWERLAW) resulted in strong residuals at $\sim$6--8 keV due to presence of an iron line and reflection hump at $\sim$15--30 keV. Two additional emission line--like features in the \textit{XMM-Newton} EPIC-pn spectra near 1.8 and 2.2 keV were also noticed. These lines most likely arise due to the calibration uncertainties near the Si and Au edges, respectively \citep{Hiemstra2011, Diaz2014}. In addition to this, a broad excess around 1 keV is clearly noticeable in Figure 7, and similar excess in EPIC-pn timing mode has been studied extensively by several authors \citep{Boirin2005, Martocchia2006, Sala2008, Hiemstra2011, Alam2015}. The reason behind the origin of this excess is not yet clear but it is thought to be related with the instrumental calibration \citep{Alam2015}. In order to further clarify this, we fitted the \textit{Swift}/XRT spectral data available on the same epochs as considered in this work with the POWERLAW and TBABS, and did not find any excess below 2.5 keV. This confirms the finding of the previous workers mentioned above that the excesses below 2.5 keV in the \textit{XMM-Newton}/EPIC-pn spectral data arise due to the calibration issues in the timing mode. We therefore excluded the EPIC-pn data below 2.5 keV in our spectral fitting.

\startlongtable
\begin{deluxetable}{cccc}
\tablecaption{Best-fit spectral parameters of the joint \textit{XMM-Newton}/EPIC-pn and \textit{NuSTAR}/FPMA, FPMB spectral data. \label{tab:table}}
\tablehead{
\colhead{Component} & \colhead{Parameter} & \colhead{Epoch - 1} & \colhead{Epoch - 2} \\
}
\startdata
		\hline
		& $\Delta{\Gamma}$ & 0.14$\pm0.01$ & 0.13$\pm0.01$ \\
		TBabs & N$_H$($\times$10$^{22} cm^{-2})$ & 2.3$^{+0.4}_{-0.3}$ & 2.40$^{+0.05}_{-0.04}$ \\
		DISKBB & kT$_{in}$(keV) & 1.1$^{+0.3}_{-0.2}$ & 1.2$\pm0.2$ \\
		& n$_{diskbb}$ & 4.70$^{+0.02}_{-2.80}$ & 3.9$^{+6.2}_{-2.0}$ \\ 
		RELXILL & i & 75$^\circ$ (f) & 75$^\circ$ (f) \\ 
			& a & 0.2 (f) & 0.2 (f) \\ 
			& q & 3(f) & 3(f) \\ 
			& r$_{in}$\tiny(ISCO) & 16.8$^{+5.9}_{-13.6}$ & 10.0$^{+3.1}_{-8.4}$\\ 
			& r$_{out}$(r$_g$) & 400 (f) & 400 (f) \\ 
			& $\Gamma$ & 1.51$^{+0.04}_{-0.05}$ & 1.51$^{+0.04}_{-0.05}$\\ 
			& A$_{Fe}$ & 3.0$^{+2.0}_{-1.0}$ & 3.0$^{+1.9}_{-0.9}$ \\ 
			& log$\varepsilon$ & 3.20$\pm0.09$ & 3.20$^{+0.04}_{-0.03}$\\ 
			& E$_{cut}$ & 92.8$^{+14.0}_{-13.8}$ & 91.9$^{+13.7}_{-13.3}$\\
			& $\mathcal{R}$ & 0.3$\pm0.1$ & 0.4$\pm$0.1\\ 
			& n$_{rel}$($\times$10$^{-3}$) & 7.2$\pm0.2$ & 6.8$\pm0.2$ \\ 
			& $\chi^2$/dof & 855.8/840 & 810.9/839 \\ 
			& $F_{abs}^p$ & 4.0 & 3.8 \\
\enddata		
\begin{tablenotes}
		\item f -- indicates the fixed parameters
		\item $p$ -- X-ray flux in the 2.5--78 keV band in units of $10^{-9}$ erg cm$^{-2}$ s$^{-1}$
		\end{tablenotes}
\end{deluxetable}

We tried to fit the iron line excess seen at $\sim$6--8 keV by adding a GAUSSIAN model component to the above mentioned model. Moreover, we also replaced the POWERLAW model by the thermally Comptonized continuum model {\scriptsize NTHCOMP}.
The model CONSTANT*$E^{\Delta{\Gamma}}$*TBABS*(DISKBB + GAUSSIAN + NTHCOMP) (hereafter Model 1) provided an acceptable fit with $\chi^2$/dof = $855.2/842$ and  $791.2/841$ for Epoch - 1 and Epoch - 2, respectively.
The centroid energy of the iron line are found to be at 6.6$\pm0.1$ keV with line width $\sigma = 0.9\pm0.1$ keV for the Epoch - 1 and 6.5$\pm0.1$ keV with line width $\sigma = 1.0\pm0.1$ keV for the Epoch - 2.
The equivalent width (EW) of the iron line are $147^{+19}_{-25}$ eV and $167.7^{+25.1}_{-33.0}$
eV for Epoch - 1 and Epoch - 2, respectively. From these calculated values, it is clear that the line energy, line width and EW of the iron line are similar within error between the two epochs.

\begin{figure}[ht!]
\plotone{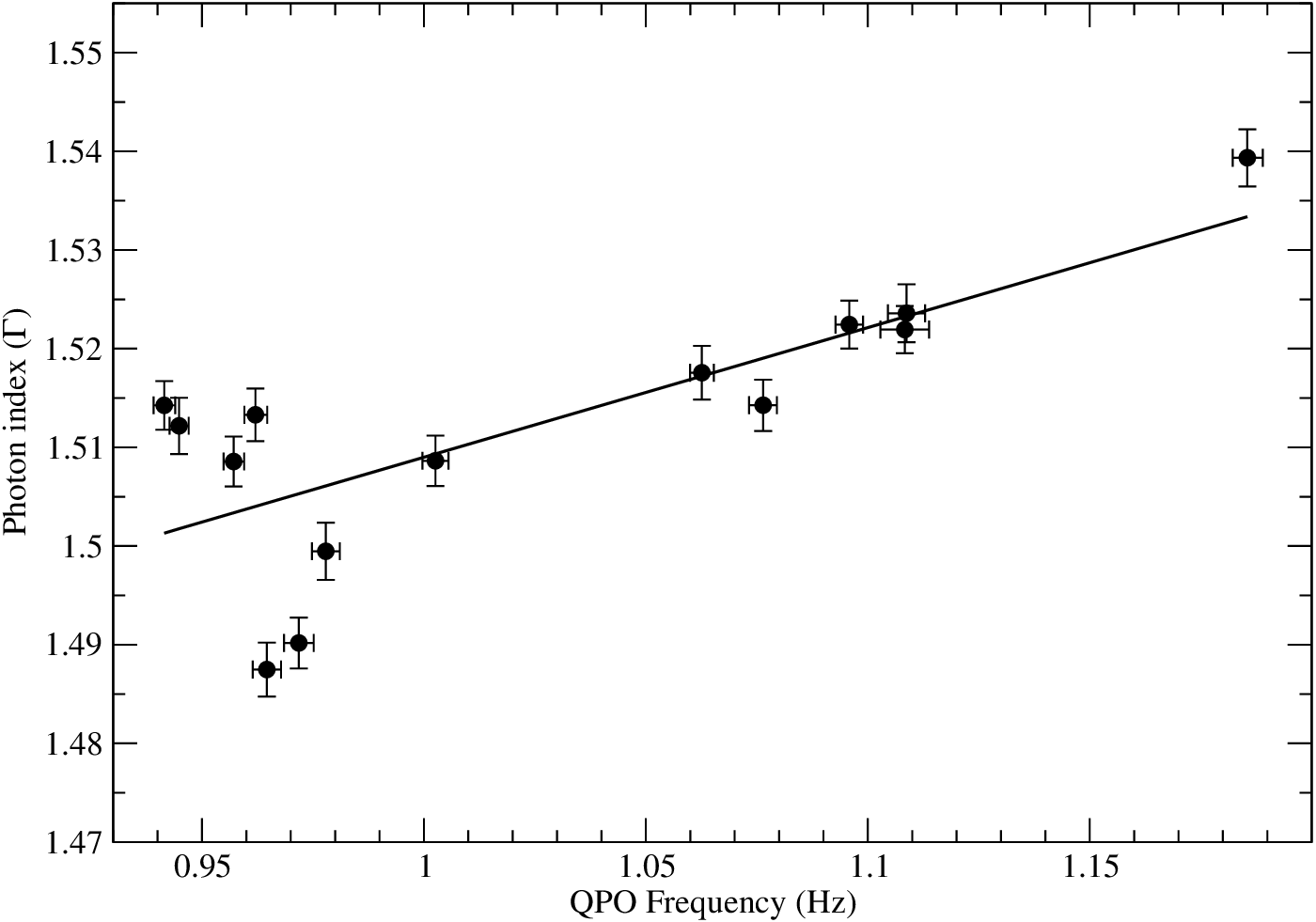}
\caption{Photon index as a function of QPO Frequency.}
\end{figure}

To fit the observed iron line and the reflection hump, we used the relativistic reflection model RELXILL \citep{Garcia et al. 2014, Dauser et al. 2014} that describes the broad iron line and reflected emission from an accretion disc illuminated by a power-law X-ray continuum with high energy cut-off.
We replaced both the GAUSSIAN and NTHCOMP components in Model 1 with RELXILL. Thus, the model CONSTANT*$E^{\Delta{\Gamma}}$*TBABS*(DISKBB+RELXILL) (hereafter Model 2)
resulted in the best fit with $\chi^2$/dof = $855.8/840$ (Epoch - 1) and $810.9/839$ (Epoch - 2).
The best-fit model (Model 2) to the \textit{XMM-Newton}/EPIC-pn and \textit{NuSTAR}
FPMA/FPMB data for both the epochs is shown in Figure 8,
whereas the corresponding best-fit parameters are listed in
Table 4.
In this, we fixed the inclination angle at $75^\circ$
and the spin parameter at 0.2 as estimated by \citet{Steiner et al. 2012}, which was also used by the previous workers for H~1743--322 \citep{Ingram and Motta 2014,
Ingram et al. 2017}. We also fixed emissivity index ($q=3$) 
for the whole disc by tying the break radius with the outermost disc radius at 400 $r_g$ \citep{Stiele and Yu 2016, Ingram et al. 2017}. 
For the absorption component, we kept the hydrogen column density parameter ($N_H$) free. In addition to this, the photon index ($\Gamma$), ionization parameter ($\xi$), as well as the iron abundance ($A_{Fe}$) were allowed to vary freely.
From the best spectral fitting, the value of the foreground absorption ($N_H$) is found to be $2.3^{+0.4}_{-0.3}$ and $2.40^{+0.05}_{-0.04}$ for the Epoch - 1 and Epoch - 2, respectively. These values are found to be almost similar to those obtained by the previous workers \citep[see][]{Stiele and Yu 2016, Parmar2003, Miller2006, CorbelTomsickKaaret2006, Shidatsu et al. 2014}.
The disc is found likely to be truncated with inner radius ($r_{in}$) 16.8$^{+5.9}_{-13.6}$ $r_{isco}$ and 10.0$^{+3.1}_{-8.4}$ $r_{isco}$ for Epoch - 1 and Epoch - 2, respectively. It is to be noted here that the truncation of the inner disc radius is not statistically significant as the uncertainties on the lower limit for the inner disc radius are large enough.
However, the inner radii are similar within errors for both the epochs.
The photon index ($\Gamma$) is $\sim$1.5 for both the epochs. The high values of the ionization parameter ($log\xi$ = $3.20\pm0.09$ and $3.20^{+0.04}_{-0.03}$ for Epoch - 1 and Epoch - 2, respectively) suggest that the disc is highly ionized.
 The reflection fraction ($\mathcal{R}$) is far
below from unity ($\sim$0.3 and $\sim$0.4 for the Epoch - 1 and Epoch - 2, respectively). Finally, the
disc temperature (kT$_{in}$) is high with the values of $\sim$1.1 and $\sim$1.2 keV for Epoch - 1 and Epoch - 2, respectively.

\begin{figure}[ht!]
\plotone{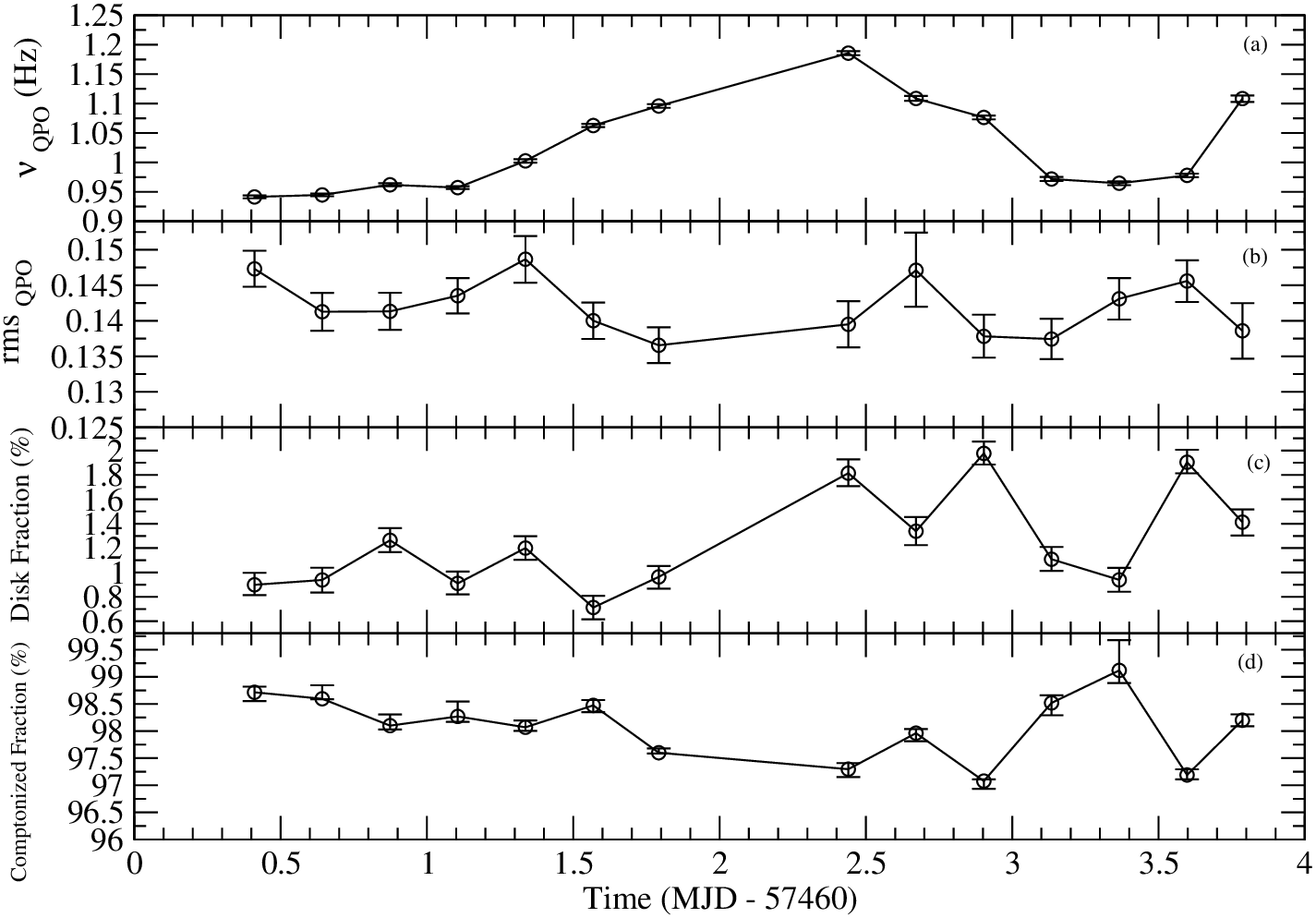}
\caption{Variation of the centroid frequency of the QPO (panel a), QPO fractional rms amplitude (panel b), disc fraction (panel c) and Comptonized fraction (panel d) with time.}
\end{figure}

\subsection{Spectral/Temporal Correlation}
To study the connections between the temporal/spectral parameters, we
divided each of the  \textit{XMM-Newton}/EPIC-pn and \textit{NuSTAR}/FPMA, FPMB datasets into seven equal 20 ks long time
intervals, giving a total number of fourteen data sets. PDSs from the corresponding background subtracted
lightcurves of the \textit{XMM-Newton} EPIC-pn data were derived, and the variation in the centroid frequency, as well as the fractional rms amplitude
of the QPO were obtained for each time interval of 20 ks. Since the significance of the upper harmonic is very low with respect to the QPO (see Section 3.1), and in the short intervals of 20 ks, the signal to noise ratio is poor, we have considered only the QPO for this part of work. In addition to this, we have also excluded the \textit{NuSTAR} data from time resolved temporal analysis due to the small number of photons in the short interval of 20 ks.
For the time resolved temporal and spectral analysis, all the errors are derived at 68\% confidence level.
From the time resolved spectral and temporal studies, we derived the photon index and the QPO frequency for each of these time intervals. Figure 9 shows the variation of the photon index ($\Gamma$) with the QPO frequency, which shows the linear correlation between these two parameters.

For each of the time-selected spectral dataset, we also computed the disc fraction and the Comptonized fraction in the 0.7--78 keV using the Model 1 by dividing the total unabsorbed flux to the disc flux and the Comptonized flux, respectively. Figure 10 shows the evolution of the spectral and temporal parameters with time, where panels (a)--(d) represent the centroid frequency of the QPO, fractional rms amplitude of the QPO, disc fraction and the Comptonized fraction, respectively. It is clear from the panels (a) and (b) that the centroid frequency and fractional rms amplitude of the QPO are weakly anti-correlated with the correlation coefficient, R $\sim$ 0.38. However, p-value of this correlation coefficient is found to be $\sim0.09$, indicating that the anti-correlation appears not to be statistically significant. This might be due to the difficulty in constraining the model parameters of the timing features as a result of poor signal to noise ratio in the short intervals of 20 ks. Moreover, the QPO frequency is weakly correlated with the disc fraction (see panels (a) and (c)), as well as weakly anti-correlated with the Comptonized fraction (see panels (a) and (d)) with the correlation coefficients, R $\sim$ 0.44 and $\sim$ 0.56, respectively. These two relations are found to be statistically significant with p-values of $\sim0.05$ and $\sim0.02$, respectively.

\section{DISCUSSION AND CONCLUDING REMARKS}

We performed the temporal and spectral analysis of H~1743--322 using the joint observations by the \textit{XMM-Newton} and \textit{NuSTAR} in two different epochs during the 2016 outburst. The HID derived from the \textit{Swift/XRT} observations (see Figure 1) indicates that the source was in the hard state during the \textit{XMM-Newton} and \textit{NuSTAR} observations, as well as the source undergoes a full spectral state transition during the 2016 outburst.
This is in contrast to the 2008 and 2014 outburst, where the source was in the hard state during the entire outburst \citep{Capitanio et al. 2009, Stiele and Yu 2016}.

We have detected QPO along with its upper harmonics at high significance levels in the 0.7--3 keV, 3--10 keV and 10--30 keV bands of \textit{XMM-Newton} and \textit{NuSTAR} data in both the epochs. The absence of the upper harmonic in the 10--30 keV band of \textit{NuSTAR} data in Epoch - 1  is due to the poor signal to noise ratio of the data.
 The shape of the PDSs, as well as the fractional rms amplitude of the QPO, derived in all the above mentioned energy bands (see Table 2 and 3), suggest that the QPO is of type C \citep{Casella et al. 2004, Motta et al. 2011}.
The \textit{NuSTAR} observations provided the opportunity to investigate the nature of the  variability of H~1743--322 in the the hard band (10--30 keV). The similar shape of the PDSs in different energy bands clearly indicates energy independent nature of the PDSs.

We noticed that the centroid frequencies of the QPO and its upper harmonic in Epoch - 2 are shifted towards higher frequency side with respect to Epoch - 1 for each energy band.
These shifts may indicate a certain geometrical change in the system between the two epochs.
In the 2010 and 2011 outbursts, \citet{Altamirano et al. 2012} reported the presence of QPO along with upper harmonic at the frequency ratio of 1:2 and a shift less than $\sim$2.2 mHz in the QPO frequency of H~1743--322 between the two successive \textit{RXTE} observations. In the failed outburst of 2014, \citet{Stiele and Yu 2016} also detected QPO and upper harmonic at the same ratio of 1:2 using a single \textit{XMM-Newton} observation.
 
On the other hand, the unchanged shape of the PDS with energy (see Figure 2 and 3) and the absence of strong energy dependence of the characteristics frequency of the QPO, as well as its upper harmonic and the zero centered BLN component (see Figure 4) are consistent with the LHS of the system observed during the rising phase of the outburst \citep{Stiele and Yu 2015, Stiele and Yu 2016}. As in Figure 5, the LHS of the system is also supported by either flat or slightly decreasing trend of the rms amplitude of the zero centered BLN with increasing  energy in both the epochs \citep{Stiele and Yu 2015, Stiele and Yu 2016}. Similar rms spectra below 10 keV were also found in few other BHXRBs such as XTE J1650--500 and XTE J1550--564 in the LHS \citep{Gierlinski and Zdziarski 2005}.

As mentioned in Section 3.3, the centroid energy, width and EW of the iron line, found during the 2016 outburst, do not change
between the two epochs, and are consistent with those found by \citet{Stiele and Yu 2016} based on the \textit{XMM-Newton} observation of the 2014 outburst. The X-ray power-law shape in both the epochs is similar ($\Gamma \sim 1.5$) and is generally found in the LHS.
Furthermore, the inner disc radius ($r_{in}$) estimated during the 2016 outburst is
found likely to be truncated away from the ISCO for both the epochs, although this is not statistically significant due to the large values of uncertainties on the lower limit (see Table 4). These results are consistent with those found by \citet{Ingram et al. 2017} during the 2014 failed outburst of H~1743--322.
However, the accretion disc
temperature pertains to be high for both epochs. We note that high inner disc
temperature with a low value of $\Gamma$ representing hard state of H~1743--322 is 
also reported by the previous workers \citep[see][]{McClintock et al. 2009,
Chen et al. 2010, Motta et al. 2011, Cheng et al. 2019}. The high disc temperature may be due to irradiation of the disc by hard X-rays from the corona. The illuminating X-rays likely get
absorbed by the disc and the disc gets thermalized, which may result in the increase of the  disc temperature \citep{Gierdonepage2009}.
The high values of ionization parameter (log$\xi\approx 3.2$) obtained in both the epochs indicate high 
 irradiation of the accretion disc by hard X-ray 
 photons from the coronal
region. As mentioned in Table 4, the energy spectra shows a high
energy cutoff value at $\sim$92 keV for both the epochs, which also indicates  the characteristic of the LHS \citep{Motta et al. 2009, Alam et al. 2014}. 
Moreover, the ratio between the
disc illuminating coronal intensity to the intensity that approaches
towards the observer is termed as the reflection fraction ($\mathcal{R}$),  which
is found to be  far below from the unity for both the epochs. Low values of the Reflection fraction may also indicate a truncated inner disc radius \citep{Garcia et al. 2015, Furst et al. 2015}, although the truncation of the inner disc radius estimated here is not statistically significant. This can also be justified by the findings of \citet{Ingram et al. 2017} who also reported the reflection fraction to be less than unity along with a truncated disc for the same source in 2014 outburst. The value of the iron line abundance is found to be high with respect to that obtained by \citet{Ingram et al. 2017}.

As shown in the bottom panels of Figure 6, time-lag-energy spectra indicates the presence of hard lag in H~1743--322 during the 2016 outburst, where the hard X-ray variations lag behind the soft X-ray variations. The hard lag is found to be $0.40\pm0.15$ s and $0.32\pm0.07$ s for the Epoch - 1 and Epoch - 2, respectively. Figure 2 indicates that the QPO does not contribute significant power in the frequency range 0.2--0.4 Hz, considered for estimating the average lags. This can also be confirmed from the top panels of Figure 6, wherein the time lag above $\sim0.4$ Hz is zero.
The presence of hard X-ray time-lag is very common for X-ray binaries. Such hard lags can be explained in terms of propagation fluctuations model \citep{Lyubarskii1997}. According to this model, the fluctuations in the mass accretion rate are different at different radii of the accretion disc and propagate down to the central object after being originated at the larger radii. As a result, the soft photons, originating from the outer part of the disc, get affected first by the fluctuation rather than the hard photons in the innermost region and thus, the hard photons lag the soft ones \citep[see][]{IngramDone2011}. Presence of hard lag can also be attributed to the delay due to the Comptonization process \citep{Uttley2014, MarcoPonti2016}. However, the lag due to the Comptonization does not completely follow the log-linear trend with energy \citep{Uttley2011}. Many previous researchers such as \citet{Pei2017}, \citet{srivivrao2009}, \citet{Sriram2007} and \citet{Choudhury2005}
 invoked the truncated disc geometry to explain the hard lag, whereas \citet{KaraErin2019} found the evidence of hard lag when the disc is consistent with being at the ISCO.
In Active Galactic Nuclei (AGN), the hard lag is commonly observed even if the accretion disk is found to be extended upto the ISCO \citep[see e.g.,][]{Kara2016, EpitropakisPapadakis2017}. Hence, the presence of hard lag in both BHXRBs and AGN is unlikely to be due to the disk truncation. It is also interesting that soft X-ray lag has been reported during the 2008 and 2014 outburst of H~1743--322 \citep{Marco2015, MarcoPonti2016}. However, these two outburst were reported to be failed ones, whereas the 2016 outburst exhibits a full spectral state transition from the hard to soft state (see Figure 1), implying to be a successful one.
As the luminosity may be the reason for the change in lag properties during the 2016 outburst from the 2008 and 2014 ones, we derived the Eddington-scaled luminosity ($L_{3-10 keV}/L_{Edd}$) in the 3-10 keV band for the 2016 outburst, considering the mass and distance to the source to be the same as that used in \citet{MarcoPonti2016}. The values of $L_{3-10 keV}/L_{Edd}$ are found to be $0.006\pm0.002$ and $0.005\pm0.002$ for Epoch - 1 and Epoch - 2, respectively. These values are nearly similar to the reported value of $L_{3-10 keV}/L_{Edd}$ $\sim0.004$ by \citet{MarcoPonti2016} in the same energy band for the 2008 and 2014 outbursts. The similar values of the Eddington-scaled luminosity indicate that there may be other possible reasons for the change in the lag properties rather than the luminosity.

 The correlation found between the QPO frequency and the photon index ($\Gamma$) (see Figure 9) is very common in BHXRBs and an ample amount of study has been performed by \citet{Vignarca2003, Titarchuk and Fiorito 2004, Titarchuk and Shaposhnikov 2005, McClintock et al. 2009, StieleMotta2011, StieleBelloni2013}. Furthermore, \citet{StieleBelloni2013} stated that the $\Gamma$-QPO frequency correlation can be explained through the `sombero' geometry, where the black hole is surrounded by the quasi-spherical corona and the accretion disc enters the corona by a little amount. Such type of $\Gamma$-QPO frequency correlation indicates that the QPO properties are strongly related to the geometry of the coronal region.

Although the weak anti-correlation between the centroid frequency and fractional rms amplitude of QPO in Figure 10 is not statistically significant, it might still give some indication of the type C nature of the QPO (McClintock et al. 2009) in addition to the obtained shape of the PDS (see Figure 2) and the fractional rms value of the QPO (see Table 2). However, there may be possibility of achieving better significance of this anti-correlation by considering the larger dataset with higher signal to noise ratio.
 The disc fraction is always less than $3\%$ and shows a weak correlation with the QPO frequency, as well as weak anti-correlation with the QPO fractional rms amplitude.
In this regard, it is worth mentioning here that based on the study of the BHXRBs 
XTE~J1550--564 and GRO~J1655--40, \citet{Sobczak et al. 2000} 
suggested that the strong correlation between the disc flux and the QPO frequency could imply the regulation of the QPO frequency 
by the accretion disc. Apart from this, it was also pointed out that not only the accretion disc regulates the QPO frequency but also 
the QPO phenomenon is closely related to the power-law component, which acts like a trigger only when the threshold 
value is $\sim20\%$ of the total flux. For H~1743--322, we have found that the Comptonized fraction is weakly anti-correlated 
with the QPO frequency, as well as weakly correlated with the QPO fractional rms amplitude 
 and is always greater than $97\%$ (see Figure 10) that indicates the maximum amount of the
 total flux is coming from the Comptonized component. The weak anti-correlation between the QPO frequency and the
 power-law flux was also found for type C QPOs in the BHXRB GX 339--4 by \citet{Motta et al. 2011}.
Moreover, the high value of the Comptonized fraction, as well as the weak value of the thermal disc component clearly support the LHS behavior of the H~1743--322 during the 2016 outburst period and is in accordance with the above mentioned temporal and spectral parameters.

\section*{Acknowledgements}

We thank the anonymous referee for useful comments that improve the quality of the paper.
This research has made use of archival data of \textit{XMM-Newton}, \textit{NuSTAR} and \textit{Swift} observatories through the High Energy Astrophysics Science Archive Research Center (HEASARC), provided by the NASA Goddard Space Flight Center. The NUSTARDAS jointly developed by the ASI Science Data Center (ASDC, Italy) and the California Institute of Technology (Caltech, USA), as well as Science Analysis System (SAS), provided by ESA science mission with instruments and contributions directly funded by ESA Member States and the USA (NASA) and \textit{Swift} online data analysis tools provided by the
Leicester Swift Data Centre (\url{http://www.swift.ac.uk/user objects/}) are used for the processing of the data of the corresponding observatories. Authors express their sincere thanks to David P. Huenemoerder for the help regarding the use of ISIS package. This research has made use of the General High-energy Aperiodic Timing Software (GHATS) package developed by T. M. Belloni at INAF--Osservatorio Astronomico diBrera. SC and PT acknowledge the financial support from grant under ISRO(AstroSat) Announcement of Opportunity (AO) programme (DS-2B-13013(2)/8/2019-Sec.2). PT expresses his sincere thanks to Inter-University Centre for Astronomy and Astrophysics (IUCAA), Pune, India for granting supports through IUCAA asscociateship programme. SC is also very much grateful to IUCAA, Pune, India for providing support and local hospitality during his frequent visits for giving the final shape to this paper.


\begin{thebibliography}{}

\bibitem[\protect\citeauthoryear{Agrawal \& Nandi}{2015}]{AgrawalNandi2015}
Agrawal, Vivek., Nandi, Anuj., 2015, MNRAS, 2015, 446, 3926
\bibitem[\protect\citeauthoryear{Alam et al.}{2014}]{Alam et al. 2014}
Alam et al., 2014, \mnras, 445, 4259
\bibitem[\protect\citeauthoryear{Alam et al.}{2015}]{Alam2015}
Alam et al., 2015, \mnras, 451, 3078
\bibitem[\protect\citeauthoryear{Altamirano et al.}{2011}]{Altamirano et al. 2011}
Altamirano D. et al., 2011, \apj, 742, L17
\bibitem[\protect\citeauthoryear{Altamirano \& Strohmayer}{2012}]{Altamirano et al. 2012}
Altamirano D., Strohmayer T., 2012, \apj, 754, L23
\bibitem[\protect\citeauthoryear{Ar{\'e}valo \& Uttley}{2006}]{ArevaloandUttley2006}
P. Ar{\'e}valo \& P. Uttley, 2006, \mnras, 367, 801
\bibitem[\protect\citeauthoryear{Bachetti}{2015}]{Bachetti2015}
Bachetti, M. 2015, MaLTPyNT: Quick look timing analysis for NuSTAR data, Astrophysics Source Code Library, ascl:1502.021
\bibitem[\protect\citeauthoryear{Belloni \& Hasinger}{1990}]{Belloni and Hasinger 1990}
Belloni T., Hasinger G., 1990, \aap, 230, 103
\bibitem[\protect\citeauthoryear{Belloni et al.}{2000}]{Belloni et al. 2000}
Belloni T., Klein-Wolt M., M ́endez M., van der Klis M., van Paradijs J., 2000, \aap, 355, 271
\bibitem[\protect\citeauthoryear{Belloni et al.}{2005}]{Belloni et al. 2005}
Belloni T., Homan J., Casella P., et al., 2005, \aap, 440, 207
\bibitem[\protect\citeauthoryear{Belloni}{2010}]{Belloni2010}
Belloni T. M., 2010, in Belloni T., ed., Lecture Notes in Physics Vol. 794, 
States and Transitions in Black Hole Binaries. Springer Verlag, Berlin, p. 53
\bibitem[\protect\citeauthoryear{Belloni, Motta \& Munoz-Darias}{2011}]{Belloni et al. 2011}
Belloni T. M., Motta S. E., Munoz-Darias T., 2011, Bulletin of the Astronomical Society of India, 39, 409
\bibitem[\protect\citeauthoryear{Belloni, Sanna \& M\'{e}ndez}{2012}]{Belloni et al. 2012}
Belloni T. M., Sanna A., M\'{e}ndez M., 2012, \mnras, 426, 1701
\bibitem[\protect\citeauthoryear{Boirin et al.}{2005}]{Boirin2005}
Boirin L., Mendez M., Da{\'i}z Trigo M., Parmar A. N., Kaastra J. S., 2005, A\&A, 436, 195
\bibitem[\protect\citeauthoryear{Burrows et al.}{2000}]{Burrows2000}
Burrows D. N. et al., 2000, in Flanagan K. A., Siegmund O. H., eds, X-Ray and Gamma-Ray Instrumentation for Astronomy XI, Vol. 4140 SPIE Conf Ser., Swift X-Ray Telescope. SPIE, Bellingham, p. 64
\bibitem[\protect\citeauthoryear{Casella et al.}{2004}]{Casella et al. 2004}
Casella P., Belloni T., Homan J., Stella L., 2004, \aap, 426, 587
\bibitem[\protect\citeauthoryear{Casella, Belloni \& Stella}{2005}]{Casella et al. 2005}
Casella, P., Belloni, T., Stella, L. 2005, \apj, 629, 403
\bibitem[\protect\citeauthoryear{Capitanio et al.}{2009}]{Capitanio et al. 2009}
Capitanio F., Belloni T., Del Santo M., Ubertini P., 2009, \mnras, 398, 1194
\bibitem[\protect\citeauthoryear{Chen et al.}{2010}]{Chen et al. 2010}
Chen, Y. P., et al., 2010, \aap, 522, A99
\bibitem[\protect\citeauthoryear{Cheng et al.}{2019}]{Cheng et al. 2019}
Cheng et al., 2019, \mnras, 482, 550
\bibitem[\protect\citeauthoryear{Choudhury et al.}{2005}]{Choudhury2005}
Choudhury, M. et al., 2005, \apj, 631, 1072
\bibitem[\protect\citeauthoryear{Corbel, Tomsick \& Kaaret}{2006}]{CorbelTomsickKaaret2006}
Corbel S., Tomsick J. A., Kaaret P., 2006, ApJ, 636, 971
\bibitem[\protect\citeauthoryear{Dauser et al.}{2010}]{Dauser et al. 2010}
Dauser, T., Wilms, J., Reynolds, C. S., and Brenneman, L. W. 2010, \mnras, 409, 1534
\bibitem[\protect\citeauthoryear{Dauser et al.}{2013}]{Dauser et al. 2013}
Dauser, T., Garc\'{i}a, J., Wilms, J., et al. 2013, \mnras, 430, 1694
\bibitem[\protect\citeauthoryear{Dauser et al.}{2014}]{Dauser et al. 2014}
Dauser, T., Garc\'{i}a, J., Parker, M. L., Fabian, A. C., and Wilms, J. 2014, \mnras, 444, L100
\bibitem[\protect\citeauthoryear{De Marco et al.}{2013}]{DeMarco2013}
De Marco, B., Ponti, G., Cappi, M., et al. 2013, MNRAS, 431, 2441
\bibitem[\protect\citeauthoryear{De Marco \& Ponti}{2016}]{MarcoPonti2016}
De Marco, B.\& Ponti, G., 2016, \apj, 826, 70
\bibitem[\protect\citeauthoryear{De Marco et al.}{2015}]{Marco2015}
De Marco, B., Ponti, G., Munoz-Darias, T., \& Nandra, K. 2015, ApJ, 814, 50
\bibitem[\protect\citeauthoryear{De Marco et al.}{2017}]{DeMarco2017}
De Marco, B. et al., 2017, MNRAS, 471, 1475
\bibitem[\protect\citeauthoryear{Dewangan, Titarchuk \& Griffiths}{2006}]{DewanganTitarchukandGriffiths2006}
Dewangan, G. C., Titarchuk, L., \& Griffiths, R. E. 2006, ApJL, 637, L21
\bibitem[\protect\citeauthoryear{D{\'i}az Trigo et al.}{2014}]{Diaz2014}
D{\'i}az Trigo et al., 2014, \aap, 571, A76
\bibitem[\protect\citeauthoryear{Epitropakis \& Papadakis}{2017}]{EpitropakisPapadakis2017}
Epitropakis A., Papadakis I.~E., 2017, MNRAS, 468, 3568
\bibitem[\protect\citeauthoryear{Esin et al.}{1997}]{Esin et al. 1997}
Esin, A. A., McClintock, J. E., and Narayan, R. 1997, \apj, 489, 865
\bibitem[\protect\citeauthoryear{Esin et al.}{2001}]{Esin et al. 2001}
Esin, A. A., McClintock, J. E., Drake, J. J., et al. 2001, \apj, 555, 483
\bibitem[\protect\citeauthoryear{Evans et al.}{2009}]{Evans2009}
Evans, P. A., Beardmore, A. P., Page, K. L., et al. 2009, \mnras, 397, 1177
\bibitem[\protect\citeauthoryear{Fabian et al.}{1989}]{Fabian et al. 1989}
Fabian A. C., Rees M. J., Stella L., White N. E., 1989, \mnras, 238, 729
\bibitem[\protect\citeauthoryear{F{\"u}rst et al.}{2015}]{Furst et al. 2015}
F{\"u}rst, F., Nowak, M. A., Tomsick, J. A., et al. 2015, \apj , 808, 122
\bibitem[\protect\citeauthoryear{Garc{\'i}a \& Kallman}{2010}]{Garcia and Kallman 2010}
Garc{\'i}a, J., Kallman, T. R. 2010, \apj, 718, 695
\bibitem[\protect\citeauthoryear{Garc{\'i}a et al.}{2011}]{Garcia et al. 2011}
Garc{\'i}a, J., Kallman, T. R., and Mushotzky, R. F. 2011, \apj, 731, 131
\bibitem[\protect\citeauthoryear{Garc{\'i}a et al.}{2013}]{Garcia et al. 2013}
Garc{\'i}a, J., Dauser, T., Reynolds, C. S., et al. 2013, \apj, 768, 146
\bibitem[\protect\citeauthoryear{Garc{\'i}a et al.}{2014}]{Garcia et al. 2014}
Garc{\'i}a J. et al., 2014, \apj, 782, 76
\bibitem[\protect\citeauthoryear{Garc{\'i}a et al.}{2015}]{Garcia et al. 2015}
Garc{\'i}a, J. A. et al., 2015, \apj, 813, 84
\bibitem[\protect\citeauthoryear{Gierli{\'n}ski \& Done}{2004}]{Gierlinski and Done 2004}
Gierli{\'n}ski, M., Done, C. 2004, \mnras, 347, 885
\bibitem[\protect\citeauthoryear{Gierli{\'n}ski \& Zdziarski}{2005}]{Gierlinski and Zdziarski 2005}
Gierli{\'n}ski M., Zdziarski A. A., 2005, \mnras, 363, 1349
\bibitem[\protect\citeauthoryear{Gierli{\'n}ski, Done \& Page}{2009}]{Gierdonepage2009}
Gierli{\'n}ski M., Done, Chris and Page, Kim, 2009, \mnras, 392, 1106
\bibitem[\protect\citeauthoryear{Grinberg et al.}{2014}]{Grinberg2014}
Grinberg, V., Pottschmidt, K., B{\"o}ck, M., et al. 2014, A\&A, 565, 1
\bibitem[\protect\citeauthoryear{Harrison et al.}{2013}]{Harrison et al. 2013}
Harrison, F.A., Craig, W.W., Christensen, F.E., et al. 2013, \apj, 770, 103
\bibitem[\protect\citeauthoryear{Hiemstra et al.}{2011}]{Hiemstra2011}
Hiemstra et al., 2011, \mnras, 411, 137
\bibitem[\protect\citeauthoryear{Hill et al.}{2000}]{Hill2000}
Hill J. E., Zugger M. E., Shoemaker J., Witherite M. E., Koch T. S., Chou L. L., Case T., Burrows D. N., 2000, in Flanagan K. A., Siegmund O. H., eds, X-Ray and Gamma-Ray Instrumentation for Astronomy XI, Vol. 4140 SPIE Conf. Ser., Laboratory X-ray CCD Camera Electronics: a Test Bed for the Swift X-Ray Telescope. SPIE, Bellingham, p. 87
\bibitem[\protect\citeauthoryear{Homan et al.}{2005}]{Homan et al. 2005}
Homan J., Miller J. M., Wijnands R., van der Klis M., Belloni T., Steeghs D., Lewin W. H. G., 2005, \apj, 623, 383
\bibitem[\protect\citeauthoryear{Houck and Denicola}{2000}]{HouckandDenicola2000}
Houck, J. C., \& Denicola, L. A. 2000, in ASP Conf. Ser. 216, Astronomical Data Analysis Software and Systems IX, ed. N. Manset, C. Veillet, \& D. Crabtree (San Francisco, CA: ASP), 591
\bibitem[\protect\citeauthoryear{Ingram \& Done}{2011}]{IngramDone2011}
Ingram A. \& Done, Chris, 2011, \mnras, 415, 2323
\bibitem[\protect\citeauthoryear{e.g. Ingram \& Motta}{2014}]{Ingram and Motta 2014}
Ingram A., Motta S., 2014, \mnras, 444, 2065
\bibitem[\protect\citeauthoryear{Ingram et al.}{2017}]{Ingram et al. 2017}
Ingram A., van der Klis. M., Middleton M. and Altamirano, D., 2017, \mnras, 464, 2979
\bibitem[\protect\citeauthoryear{Kaluzienski \& Holt}{1977}]{Kaluzienski and Holt 1977}
Kaluzienski L. J., Holt S. S., 1977, IAU Circ., 3099, 1
\bibitem[\protect\citeauthoryear{Kara et al.}{2016}]{Kara2016}
Kara et al., 2016, \mnras, 462, 511
\bibitem[\protect\citeauthoryear{Kara et al.}{2019}]{KaraErin2019}
Kara E. et al., 2019, Nature, 565, 198
\bibitem[\protect\citeauthoryear{Krimm et al.}{2009}]{Krimm et al. 2009}
Krimm H. A. et al., 2009, Astron. Telegram, 2058
\bibitem[\protect\citeauthoryear{Leahy et al.}{1983}]{Leahyetal1983}
Leahy, D. A., Elsner, R. F., \& Weisskopf, M. C. 1983, ApJ, 272, 256
\bibitem[\protect\citeauthoryear{Lyubarskii}{1997}]{Lyubarskii1997}
Lyubarskii, Yu. E., 1997, \mnras, 292, 679
\bibitem[\protect\citeauthoryear{Madsen et al.}{2017}]{Madsen2017}
Madsen, K. K. et al., 2017, \aj, 153, 2
\bibitem[\protect\citeauthoryear{Markwardt \& Swank}{2003}]{Markwardt and Swank 2003}
Markwardt C. B., Swank J. H., 2003, Astron. Telegram, 133
\bibitem[\protect\citeauthoryear{Martocchia et al.}{2006}]{Martocchia2006}
Martocchia A., Matt G., Belloni T., Feroci M., Karas V., Ponti G., 2006, A\&A, 448, 677
\bibitem[\protect\citeauthoryear{McClintock, Horne \& Remillard}{1995}]{McClintock et al. 1995}
McClintock, J. E., Horne, K., and Remillard, R. A. 1995, \apj, 442, 358
\bibitem[\protect\citeauthoryear{McClintock et al.}{2001}]{McClintock et al. 2001}
McClintock, J. E., Haswell, C. A., Garcia, M. R., et al. 2001, \apj, 555, 477
\bibitem[\protect\citeauthoryear{McClintock et al.}{2003}]{McClintock et al. 2003}
McClintock, J. E., Narayan, R., Garcia, M. R., et al. 2003, \apj, 593, 435
\bibitem[\protect\citeauthoryear{McClintock \& Remillard}{2006}]{McClintock and Remillard 2006}
McClintock J. E., Remillard R. A., 2006, in Compact Stellar X-ray Sources, vol. 39, ed. W. Lewin and M. van der Klis (Cambridge: Cambridge univ. Press), 157
\bibitem[\protect\citeauthoryear{McClintock et al.}{2009}]{McClintock et al. 2009}
McClintock, J. E., Remillard, R. A., Rupen, M. P., et al. 2009, \apj, 698, 1398
\bibitem[\protect\citeauthoryear{Miyamoto et al.}{1988}]{Miyamoto1988}
Miyamoto, S., Kitamoto, S., Mitsuda, K., \& Dotani, T. 1988, Natur, 336, 450
\bibitem[\protect\citeauthoryear{Miller et al.}{2006}]{Miller2006}
Miller J. M. et al., 2006, ApJ, 646, 394
\bibitem[\protect\citeauthoryear{Miller}{2007}]{Miller 2007}
Miller J. M., 2007, \araa, 45, 441
\bibitem[\protect\citeauthoryear{Mitsuda et al.}{1984}]{Mitsuda et al. 1984}
Mitsuda, K., Inoue, H., Koyama, K., et al. 1984, \pasj, 36, 741
\bibitem[\protect\citeauthoryear{Molla et al.}{2017}]{Molla et al. 2017}
Molla, A. A., et al., 2017, \apj, 834, 88
\bibitem[\protect\citeauthoryear{Morgan, Remillard \& Greiner}{1997}]{Morgan et al. 1997}
Morgan E. H., Remillard R. A., Greiner J., 1997, \apj, 482, 993
\bibitem[\protect\citeauthoryear{Motta, Belloni \& Homan}{2009}]{Motta et al. 2009}
Motta S., Belloni T., Homan J., 2009, \mnras, 400, 1603
\bibitem[\protect\citeauthoryear{Motta, Munoz-Darias \& Belloni}{2010}]{Motta et al. 2010}
Motta S., Munoz-Darias T., Belloni T., 2010, \mnras, 408, 1796
\bibitem[\protect\citeauthoryear{Motta et al.}{2011}]{Motta et al. 2011}
Motta S., Munoz-Darias T., Casella P., Belloni T., Homan J., 2011, \mnras, 418, 2292
\bibitem[\protect\citeauthoryear{Motta et al.}{2017}]{Motta et al. 2017}
Motta S. E., Rouco-Escorial A., Kuulkers E., Munoz-Darias T., Sanna A., 2017, \mnras, 468, 2311
\bibitem[\protect\citeauthoryear{Nakahira et al.}{2013}]{Nakahira et al. 2013}
Nakahira S. et al., 2013, Astron. Teleg., 5241, 1
\bibitem[\protect\citeauthoryear{Narayan and Yi}{1995}]{Narayan and Yi 1995}
Narayan, R., Yi, I. 1995, \apj, 452, 710
\bibitem[\protect\citeauthoryear{Narayan, McClintock \& Yi}{1996}]{Narayan et al. 1996}
Narayan, R., McClintock, J. E., and Yi, I. 1996, \apj, 457, 821
\bibitem[\protect\citeauthoryear{Negoro et al.}{2012}]{Negoro et al. 2012}
Negoro H. et al., 2012, Astron. Teleg., 3842, 1
\bibitem[\protect\citeauthoryear{Nolan et al.}{1981}]{Nolan1981}
Nolan P. L. et al., 1981, ApJ, 246, 494
\bibitem[\protect\citeauthoryear{Nowak et al.}{1999a}]{Nowak1999a}
Nowak, M. A., Vaughan, B. A., Wilms, J., Dove, J. B., \& Begelman, M. C. 1999a, ApJ, 510, 874
\bibitem[\protect\citeauthoryear{Nowak et al.}{1999b}]{Nowak1999b}
Nowak, M. A., Wilms, J., \& Dove, J. B. 1999b, ApJ, 517, 355
\bibitem[\protect\citeauthoryear{Page et al.}{1981}]{Page1981}
Page, C. G., Bennetts, A. J., \& Ricketts, M. J. 1981, SSRv, 30, 369
\bibitem[\protect\citeauthoryear{Parmar et al.}{2003}]{Parmar2003}
Parmar A. N., Kuulkers E., Oosterbroek T., Barr P., Much R., Orr A., Williams O. R., Winkler C., 2003, A\&A, 411, L421
\bibitem[\protect\citeauthoryear{Pei et al.}{2017}]{Pei2017}
Pei, Songpeng. et al., 2017, Ap\&SS, 362, 118
\bibitem[\protect\citeauthoryear{Plant et al.}{2015}]{Plant et al. 2015}
Plant, D. S. et al., 2015, \aap, 573, 120
\bibitem[\protect\citeauthoryear{Priedhorsky et al.}{1979}]{Priedhorsky1979}
Priedhorsky W., Garmire G. P., Rothschild R., Boldt E., Serlemitsos P., Holt S., 1979, ApJ, 233, 350
\bibitem[\protect\citeauthoryear{Remillard et al.}{2006}]{Remillard et al. 2006}
Remillard, R. A., McClintock, J. E., Orosz, J. A., Levine, A. M. 2006, ApJ, 637, 1002
\bibitem[\protect\citeauthoryear{Revnivstev}{2003}]{Revnivstev 2003}
Revnivtsev M., 2003, \aap, 410, 865
\bibitem[\protect\citeauthoryear{Reynolds \& Nowak}{2003}]{Reynolds and Nowak 2003}
Reynolds C. S., Nowak M. A., 2003, \physrep, 377, 389
\bibitem[\protect\citeauthoryear{Sala et al.}{2008}]{Sala2008}
Sala G., Greiner J., Ajello M., Primak N., 2008, A\&A, 489, 1239
\bibitem[\protect\citeauthoryear{Shakura \& Sunyaev}{1973}]{Shakura and Sunyaev 1973}
Shakura, N. I., Sunyaev, R. A. 1973, \aap, 24, 337
\bibitem[\protect\citeauthoryear{Shaposhnikov \& Titarchuk}{2007}]{Shaposhnikov and Titarchuk 2007}
Shaposhnikov, N., Titarchuk, L. 2007, \apj, 663, 445
\bibitem[\protect\citeauthoryear{Shidatsu et al.}{2012}]{Shidatsu et al. 2012}
Shidatsu, M., Negoro, H., Nakahira, S., et al. 2012, ATel, 4419, 1
\bibitem[\protect\citeauthoryear{Shidatsu et al.}{2014}]{Shidatsu et al. 2014}
Shidatsu M. et al. 2014, \apj, 789, 100
\bibitem[\protect\citeauthoryear{Sobczak et al.}{2000}]{Sobczak et al. 2000}
Sobczak, G. J. et al., 2000, \apj, 531, 537
\bibitem[\protect\citeauthoryear{Sriram et al.}{2007}]{Sriram2007}
Sriram, K., Agrawal, V.K., Pendharkar, J.K., Rao, A.R., 2007, \apj, 661, 1055
\bibitem[\protect\citeauthoryear{Sriram, Agrawal \& Rao}{2009}]{srivivrao2009}
Sriram, K., Agrawal, V. K., Rao, A. R., 2009, RAA, 9, 901
\bibitem[\protect\citeauthoryear{Steiner et al.}{2010}]{Steiner et al. 2010}
Steiner, J. F., McClintock, J. E., Remillard, R. A., et al. 2010, \apj, 718, L117
\bibitem[\protect\citeauthoryear{Steiner et al.}{2012}]{Steiner et al. 2012}
Steiner F. J., McClintock J. E., Reid M. J., 2012, \apj, 745, 7
\bibitem[\protect\citeauthoryear{Stiele et al.}{2011}]{StieleMotta2011}
Stiele, H., Motta, S., Muñoz-Darias, T. \& Belloni, T. M., 2011, \mnras, 418, 1746
\bibitem[\protect\citeauthoryear{Stiele et al.}{2013}]{StieleBelloni2013}
Stiele, H., Belloni, T. M., Kalemci, E. \& Motta, S., 2013, \mnras, 429, 2655
\bibitem[\protect\citeauthoryear{Stiele \& Yu}{2015}]{Stiele and Yu 2015}
Stiele, H., Yu W., 2015, \mnras, 452, 3666
\bibitem[\protect\citeauthoryear{Stiele \& Yu}{2016}]{Stiele and Yu 2016}
Stiele, H., Yu, W., 2016, \mnras, 460, 1946
\bibitem[\protect\citeauthoryear{Stiele \& Kong}{2017}]{Stiele and Kong 2017}
Stiele, H., Kong, A. K. H., 2017, \apj, 844, 8
\bibitem[\protect\citeauthoryear{Tanaka \& Shibazaki}{1996}]{Tanaka and Shibazaki 1996}
Tanaka y., Shibazaki N., 1996, \araa, 34, 607
\bibitem[\protect\citeauthoryear{Titarchuk \& Fiorito}{2004}]{Titarchuk and Fiorito 2004}
Titarchuk, L., Fiorito, R. 2004, \apj, 612, 988
\bibitem[\protect\citeauthoryear{Titarchuk \& Shaposhnikov}{2005}]{Titarchuk and Shaposhnikov 2005}
Titarchuk, L., \& Shaposhnikov, N. 2005, \apj, 626, 298
\bibitem[\protect\citeauthoryear{Trudolyubov, Borozdin \& Priedhorsky}{2001}]{Trudolyubov et al. 2001}
Trudolyubov S. P., Borozdin K. N., Priedhorsky W. C., 2001, \mnras, 322, 309
\bibitem[\protect\citeauthoryear{Uttley et al.}{2011}]{Uttley2011}
Uttley P., Wilkinson T., Cassatella P., Wilms J., Pottschmidt K., Hanke M., Böck M., 2011, \mnras, 414, 60
\bibitem[\protect\citeauthoryear{Uttley et al.}{2014}]{Uttley2014}
Uttley, P., Cackett, E. M., Fabian, A. C., Kara, E., \& Wilkins, D. R. 2014, A\&ARv, 22, 72
\bibitem[\protect\citeauthoryear{Verner et al.}{1996}]{Verner et al. 1996}
Verner, D. A., Ferland, G. J., Korista, K. T., and Yakovlev, D. G. 1996, \apj, 465, 487
\bibitem[\protect\citeauthoryear{Vignarca et al.}{2003}]{Vignarca2003}
Vignarca, F., Migliari, S., Belloni, T., Psaltis, D., \& van der Klis, M. 2003, \aap, 
397, 729
\bibitem[\protect\citeauthoryear{Wilms, Allen \& McCray}{2000}]{Wilms et al. 2000}
Wilms, J., Allen, A., and McCray, R. 2000, \apj, 542, 914
\bibitem[\protect\citeauthoryear{Zhang et al.}{1995}]{Zhangetal1995}
Zhang, W., Jahoda, K., Swank, J. H., Morgan, E. H., \& Giles, A. B. 1995, ApJ, 449, 930
\bibitem[\protect\citeauthoryear{Zhou et al.}{2013}]{Zhou et al. 2013}
Zhou, J. N., Liu, Q. Z., et al., 2013, \mnras, 431, 2285  

\end{thebibliography}
\end{document}